\documentclass[a4paper,11pt]{article}
\usepackage{jheppub}
\usepackage{booktabs}
\usepackage{array}
\usepackage{color}
\usepackage{tikz}
\usetikzlibrary{decorations.markings,arrows.meta}

\title{Massive Type IIA Forms and the Romans Mass from a $\mathrm{GL}(1,\mathbb{R})$ Topological Field Theory with Boundary}

\author[a]{Euihun Joung}
\author[a]{and J.~A.~Rosabal}
\affiliation[a]{Department of Physics and Research Institute of Basic Science, Kyung Hee University,\\ Seoul 02447, Republic of Korea}
\emailAdd{euihun.joung@khu.ac.kr}
\emailAdd{jarosabal80@khu.ac.kr}

\abstract{We construct an eleven-dimensional topological field theory with boundary
whose edge modes are the dilaton and form fields of type IIA supergravity, including
the Romans mass. The bulk action is the Chern--Simons functional of the superalgebra
of the doubled formalism, completed by two brackets that are invisible in ten
dimensions; it is unique up to normalization, contains no potentials, and the metric
enters only through a boundary term. The Romans mass is the boundary value of a
zero-form, like the other field strengths: it is varied, and its equation of motion
is an eleven-form Gauss law with no ten-dimensional counterpart. The boundary term
breaks the $GL(1,\mathbb{R})$ gauge symmetry to $O(1)$: the dilaton is the edge mode
on $GL(1,\mathbb{R})/O(1)$, and $(-1)^{F_L}$, which reverses the Romans mass, becomes
a gauge symmetry. The holonomies of the flat connection are monodromies of the dilaton
and of $(-1)^{F_L}$, fixed classically by the D8-brane charges when the mass is
nonzero, and the hypersurface dual to the $O(1)$ holonomy can end on reflection
seven-branes. Since no field is self-dual, the reduction gives Romans' action off
shell, while other boundary conditions of the same bulk give massless type IIA and
dual formulations.}

\begin{document}
\maketitle
\flushbottom

\section{Introduction}

Massive type IIA supergravity~\cite{Romans:1985tz} differs from the other ten-dimensional supergravities by a mass parameter $m$. In the supergravity action, $m$ is a constant, fixed once and for all. In string theory, it is the flux of a zero-form RR field strength $F_0$: it is quantized, and it jumps across D8-branes~\cite{Polchinski:1995mt,Bergshoeff:1995super8,Bergshoeff:1996ui}. A description that holds on both sides of a D8-brane must therefore treat $m$ as a field. This requires a nine-form potential~\cite{Bergshoeff:1995super8}, and leads to the democratic formulation~\cite{Bergshoeff:2001pv}, in which every RR field strength, including $F_0$, appears together with its dual. Its action, however, is a pseudo-action: the duality relations are not among its equations of motion, and must be imposed by hand. A genuine duality-symmetric action that includes the Romans mass, the dilaton and their duals was constructed in the PST formalism~\cite{Bandos:2003et}, at the price of an auxiliary scalar that enters non-polynomially, as in the analogous type IIB action~\cite{DallAgata:1998va}. Democratic actions with auxiliary fields~\cite{Mkrtchyan:2022xrm} and democratic pseudo-actions that include the scalars~\cite{FernandezMelgarejo:2023kwk} have also been given. Polynomial, manifestly Lorentz-invariant actions for self-dual and duality-symmetric fields have been given in Sen's formulation~\cite{Sen:2015nph,Sen:2019qit}, and, in terms of potentials, in flat~\cite{Chakrabarti:2025manifest} and curved~\cite{Chakrabarti:2026manifest} spacetime.

A different route is to regard the ten-dimensional fields as the boundary values of a topological field theory in one higher dimension. The idea originates in the study of chiral scalars~\cite{Floreanini:1987as} and of chiral or duality-invariant form fields~\cite{Henneaux:1988gg,Schwarz:1993vs}, which are most naturally defined through a topological bulk~\cite{Witten:1996hc,Witten:1999vg,Maldacena:2001ss,Hopkins:2002rd,Belov:2006jd,Freed:2012bs,Heckman:2017uxe,Arvanitakis:2022bnr,Chen:2025higher}. In the bulk, a form and its dual are canonically conjugate, no metric is needed, and the duality relation becomes a boundary condition, which a metric-dependent boundary term makes follow from the variational principle~\cite{Maldacena:2001ss,Arvanitakis:2022bnr,Evnin:2023ypu}. An early example is Ref.~\cite{Lavrinenko:1999bulk}, which related higher-rank Chern--Simons theories in twelve and eleven dimensions to the form-field sectors of eleven-dimensional supergravity and of massless type IIA and type IIB, with the scalars entering through their derivatives; there, the duality relations are imposed as boundary conditions, without a boundary term. Genuinely democratic actions have since been obtained for the form fields of eleven-dimensional supergravity~\cite{Evnin:2023ypu,Rosabal:2025} and for the RR fields of type II supergravities, either on a fixed NSNS background~\cite{Belov:2006xj}, or with the NSNS fields, the dilaton and gravity added non-democratically~\cite{Evnin:2023ypu}. Eleven-dimensional topological theories of the type IIA fields also appear in the SymTFT literature~\cite{Apruzzi:2023uma,GarciaEtxebarria:2024descent}: in the symmetry descent of Ref.~\cite{GarciaEtxebarria:2024descent}, $H_3$ is a classical background, and in M-theory, this role is played by a twelve-dimensional BF theory with a cubic term and a metric-dependent boundary term~\cite{Banerjee:2026sym}. Scalars fit into the same framework: scalars on a coset $G/H$ are the edge modes of a topological gauge theory with gauge group $G$, broken to $H$ on the boundary~\cite{Moore:1989yh,Bonetti:2024symtft,Antinucci:2024ssb,Apruzzi:2025spacetime,Bonetti:2025nonlinear}.

In Ref.~\cite{Joung:IIB}, we combined these two constructions for type IIB supergravity. The theory is defined on an eleven-manifold $M_{11}$ with boundary $M_{10}$. Its fields are forms $A_p$, one for each multiplet of field strengths of democratic type IIB, of the same degree and in the same representation of $\mathrm{SL}(2,\mathbb{R})$. The one-form $A_1$ is an $\mathrm{SL}(2,\mathbb{R})$ connection, which the equations of motion require to be flat, and the other forms couple to it; in Ref.~\cite{Lavrinenko:1999bulk}, the scalars enter instead through the two one-forms ${\rm d}\phi$ and $e^\phi{\rm d}\chi$. The bulk action is topological, and its equations of motion, the Gauss laws, reduce on $M_{10}$ to the Bianchi identities. A boundary term, the only place where the metric enters, breaks $\mathrm{SL}(2,\mathbb{R})$ to $\mathrm{SO}(2)$. The axion--dilaton then arises as the edge mode on $\mathrm{SL}(2,\mathbb{R})/\mathrm{SO}(2)$, and the boundary equations of motion are the duality relations, including the self-duality of the five-form. Since the action involves the flat connection and not the axion--dilaton itself, S-duality monodromies are holonomies. The five-form, being its own dual, cannot be eliminated off shell, so that the reduction to the boundary gives the type IIB pseudo-action only after solving the Gauss laws of the bulk, including that of the five-form itself.

Here we carry out this construction for massive type IIA. The gauge group is now abelian, $G=\mathrm{GL}(1,\mathbb{R})\cong\mathbb{R}^+\times\mathrm{O}(1)$: the factor $\mathbb{R}^+$ is the scaling symmetry that shifts the dilaton, and $\mathrm{O}(1)\cong\mathbb{Z}_2$ will turn out to be generated by $(-1)^{F_L}$. Our main result is the action $S_{\rm bTFT}=S_{\rm TFT}+S_{\rm b}$ of Eqs.~\eqref{eq:S11}--\eqref{eq:S10}. Its fields are forms $A_p$, one for each field strength of democratic type IIA and of the same degree. They include a flat $\mathrm{GL}(1,\mathbb{R})$ connection $A_1$ for the dilaton and a zero-form $A_0$ for the Romans mass, whose constancy is a Gauss law. The bulk term $S_{\rm TFT}$ is topological, polynomial, free of auxiliary fields and potentials, and fixed by the gauge algebra. The boundary term $S_{\rm b}$ breaks $\mathrm{GL}(1,\mathbb{R})$ to $\mathrm{O}(1)$, so that the dilaton is the edge mode on $\mathrm{GL}(1,\mathbb{R})/\mathrm{O}(1)\cong\mathbb{R}^+$. This term is an input: it is assumed to be quadratic, and three combinations of its coefficients are chosen so as to reproduce type IIA, where supersymmetry fixes them (Sec.~\ref{sec:bdry}). What follows from the bulk alone are the ratios $\frac52:\frac32:\frac12:-1$ of the dilaton couplings of $F_0$, $F_2$, $F_4$ and $H_3$, the St\"uckelberg couplings, and the relative coefficients of the Chern--Simons terms.

The construction continues the approach of Refs.~\cite{Evnin:2023ypu,Joung:IIB}, and rests on three features. First, as in Ref.~\cite{Evnin:2023ypu}, each form and its dual enter the bulk term symmetrically, and the boundary term treats them alike; it is this choice that makes the boundary conditions the duality relations. Second, as in Ref.~\cite{Joung:IIB}, a one-form is part of the bulk: it is a flat connection, under which the other forms are charged, and the scalars arise from it as edge modes when the boundary term breaks the gauge symmetry. This non-abelian connection and its breaking were the key ingredients of Ref.~\cite{Joung:IIB}. Third, and this is the key new ingredient here, the zero-form $A_0$, which has no counterpart in type IIB or in Ref.~\cite{Lavrinenko:1999bulk}, is a field like the others. Varying it gives a Gauss law of degree eleven, which has no ten-dimensional counterpart. Its integral over the bulk is the condition that the action be stationary with respect to the Romans mass; for nonzero Romans mass, this condition already follows from the other field equations (Sec.~\ref{sec:zeromode}). Being a zero-form, $A_0$ could in principle enter non-polynomially; the action considered here is cubic. The zero-form matters beyond ten dimensions. In lower-dimensional supergravities, the deformation parameters of a gauging sit at the same degree, paired with top forms, and the gauging produces a scalar potential; in ten dimensions, $A_0$ already gives the potential term of Romans' theory. Lower-dimensional maximal supergravities will be treated along these lines in a forthcoming paper~\cite{Joung:lower}. The construction is still limited in scope: the metric remains a fixed background, fermions are not included, and the one-loop coupling $B_2\wedge X_8$~\cite{Vafa:1995fj,Duff:1995wd}, tied to the anomaly of the NS5-brane, is yet to be incorporated. The eleven-dimensional topological action of Ref.~\cite{Apruzzi:2023uma} includes it through a term $H_3\wedge X_8$, but in our action such a term would break the $\mathbb{R}^+$ gauge symmetry, as expected for a one-loop correction (Sec.~\ref{sec:democratic}).

The paper is organized as follows. Section~\ref{sec:bulk} presents the action, its local $\mathrm{GL}(1,\mathbb{R})$ symmetry, its Gauss laws, and the boundary conditions implied by the boundary term. Section~\ref{sec:reduction} carries out the reduction to the boundary. Section~\ref{sec:romans} identifies the result with Romans' action, whose Chern--Simons terms, including the $B_2^5$ term, descend from a single cubic term of the bulk~\cite{Bergshoeff:1995super8}. It also relates the full set of equations to the democratic formulation, shows that $\mathrm{O}(1)$ acts as $(-1)^{F_L}$, identifies an orientation-reversing symmetry related to the orientifold reflection $I_9\Omega$, and derives the equation of motion of the Romans mass from the eleven-form Gauss law. Section~\ref{sec:global} describes the global structure: the holonomies of the flat connection are monodromies of the dilaton and of $(-1)^{F_L}$, and for nonzero Romans mass they are fixed, classically, by the D8-brane charges; the hypersurface dual to the $\mathrm O(1)$ holonomy can end on reflection seven-branes. Branes are treated in the eleven-dimensional bulk only as far as the holonomies require; a general treatment will be given in a forthcoming paper~\cite{Joung:inprep}. Section~\ref{sec:algebra} identifies the algebra behind the construction. The bulk fields span the superalgebra of the doubled formalism~\cite{Cremmer:1998px,Lavrinenko:1999xi}, completed by two brackets that are invisible in ten dimensions, and the bulk term is the Chern--Simons functional built on its invariant pairing. Both brackets enter the Gauss law of degree eleven, and one of them relates the D8-brane tension to those of the D6-brane and the fundamental string~\cite{Lavrinenko:1999xi}. The boundary term is written in the same language, in terms of a linear map on the superalgebra, the twist, whose form is fixed up to three couplings. Topological boundary conditions of the same bulk give dual formulations, and imposing one on some of the fields and the twist on the others gives, for instance, massless type IIA. Finally, the superalgebra is compared with Borcherds superalgebras~\cite{HenryLabordere:2002dk,Greitz:2011vh} and tensor hierarchy algebras~\cite{Palmkvist:2013vya}. Section~\ref{sec:discussion} summarizes the results, compares them with type IIB, and lists the open questions. Appendix~\ref{app:conventions} collects our conventions and the Lorentzian form of the action.

\section{The topological field theory}
\label{sec:bulk}

We work on an eleven-manifold $M_{11}$ with boundary $\partial M_{11}=M_{10}$. The fields are ten differential forms $A_p$ on $M_{11}$, one in each degree $p=0,\dots,10$ except five, and the action is $S_{\rm bTFT}=S_{\rm TFT}+S_{\rm b}$, with
\begin{align}
S_{\rm TFT}&=\frac i2\int_{M_{11}}\Big\{2\,\big(A_1\,{\rm d} A_9+A_9\,{\rm d} A_1\big)+A_3\,{\rm D} A_7+A_7\,{\rm D} A_3
\nonumber\\
&\hspace{4.5em}+\sum_{k=0}^{5}(-1)^k\,A_{2k}\big({\rm D} A_{10-2k}-A_3A_{8-2k}\big)\Big\}\,,
\label{eq:S11}\\[4pt]
S_{\rm b}&=\frac{1}{4}\int_{M_{10}}\Big\{4\,A_1*A_1+A_9*A_9+A_3*A_3+A_7*A_7+\sum_{k=0}^{5}A_{2k}*A_{2k}\Big\}\,,
\label{eq:S10}
\end{align}
where wedge products are implicit, $A_{-2}\equiv0$,\footnote{Throughout, the Einstein-frame metric is a fixed background, we set $2\kappa_{10}^2=1$, and fermions are not considered. We work in Euclidean signature with complexified fields, so that $*^2=-1$ on odd forms and $*^2=+1$ on even forms in ten dimensions; all relations become real in Lorentzian signature. Our conventions, the real slice of the Euclidean fields and the Lorentzian action are given in Appendix~\ref{app:conventions}.} and
\begin{equation}
{\rm D} A_p={\rm d} A_p-w_p\,A_1A_p
\label{eq:covD}
\end{equation}
is a covariant derivative with respect to the one-form $A_1$, with the weights
\begin{equation}
\arraycolsep=6pt\def\arraystretch{1.4}\begin{array}{c|cccccccccc}
p & 0 & 1 & 2 & 3 & 4 & 6 & 7 & 8 & 9 & 10\\\hline
w_p & \frac52 & 0 & \frac32 & -1 & \frac12 & -\frac12 & 1 & -\frac32 & 0 & -\frac52
\end{array}
\label{eq:weights}
\end{equation}
The bulk term $S_{\rm TFT}$ is polynomial, of first order in derivatives and free of auxiliary fields, and the metric enters only through the boundary term $S_{\rm b}$. In Sec.~\ref{sec:romans}, the boundary values of the $A_p$ will become the field strengths of democratic massive type IIA, with $A_0$ proportional to the Romans mass and $A_1$ to the derivative of the dilaton.  We first describe the bulk term, its gauge symmetry and its equations of motion (Sec.~\ref{sec:weights}), and then the boundary term and the boundary conditions that it implies (Sec.~\ref{sec:bdry}).

\subsection{Bulk action, gauge symmetry and Gauss laws}
\label{sec:weights}

The bulk term $S_{\rm TFT}$ consists of kinetic terms and cubic interactions. The kinetic terms pair each $A_p$ with $A_{10-p}$: in eleven dimensions, a $p$-form and a $(10-p)$-form are canonically conjugate through $\int A_p\,{\rm d} A_{10-p}$. We refer to $(A_p,A_{10-p})$ as a symplectic pair. Since degree five is empty, no field is paired with itself, and the degree-zero field $A_0$ is paired with $A_{10}$. Each pair enters in both orders, which agree up to a total derivative; for the even forms, the alternating sign $(-1)^k$ compensates the sign produced by the exchange. The interactions are the cubic couplings $A_3A_{2k}A_{8-2k}$ of $A_3$ to the even forms, and the couplings to $A_1$ contained in ${\rm D}$.

The one-form $A_1$ is the connection of a local $\mathrm{GL}(1,\mathbb{R})$ symmetry. The gauge group is the product of its identity component and its maximal compact subgroup,
\begin{equation}
\mathrm{GL}(1,\mathbb{R})\cong\mathbb{R}^+\times\mathrm{O}(1)\,,\qquad \lambda=\sigma\,e^{\varepsilon_0/2}\,,\quad \sigma=\mathrm{sgn}\,\lambda=\pm1\,,
\label{eq:split}
\end{equation}
and its one-dimensional real representations are therefore the characters $\lambda\mapsto|\lambda|^{s}\sigma^{\eta}$, with $s\in\mathbb R$ and $\eta\in\{0,1\}$~\cite{Tate:1967}. Accordingly, each field carries two labels: a weight $w_p$, which fixes the action of $\mathbb{R}^+$, and a parity $\eta_p$, which fixes the action of $\mathrm{O}(1)$. A gauge transformation with parameter $\lambda$ acts as
\begin{equation}
A_1\to A_1+{\rm d}\varepsilon_0\,,\qquad A_p\to e^{w_p\varepsilon_0}\,\sigma^{\eta_p}A_p\quad(p\neq1)\,,
\label{eq:Rgauge}
\end{equation}
with $e^{w_p\varepsilon_0}=|\lambda|^{2w_p}$ and ${\rm d}\varepsilon_0=2\,\lambda^{-1}{\rm d}\lambda$. The factors of two are a matter of normalization, on which we comment below.

Besides \eqref{eq:Rgauge}, the bulk term has higher-form gauge symmetries, with a $(p-1)$-form parameter $\varepsilon_{p-1}$ of weight $w_p$ for each $A_p$ with $p\geq2$:
\begin{align}
\delta A_3&={\rm D}\varepsilon_2\,,\nonumber\\
\delta A_{2k}&={\rm D}\varepsilon_{2k-1}+A_{2k-2}\,\varepsilon_2-A_3\,\varepsilon_{2k-3}\,,\qquad k=0,\dots,5\,,\nonumber\\
\delta A_7&={\rm D}\varepsilon_6+\sum_{k=0}^{3}(-1)^k\,A_{2k}\,\varepsilon_{7-2k}\,,\nonumber\\
\delta A_9&={\rm d}\varepsilon_8+\tfrac12\big(A_3\,\varepsilon_6-A_7\,\varepsilon_2\big)-\tfrac12\sum_{k=0}^{4}(-1)^k\,w_{2k}\,A_{2k}\,\varepsilon_{9-2k}\,,
\label{eq:hfgauge}
\end{align}
where ${\rm D}$ acts on $\varepsilon_{p-1}$ with the weight $w_p$, $\varepsilon_{q}\equiv0$ for $q<0$, and $A_1$ and $A_0$ are invariant. Like \eqref{eq:Rgauge}, they leave $S_{\rm TFT}$ invariant up to boundary terms, and they are reducible, as usual for higher-form gauge fields. Together with the $\mathbb{R}^+$ transformations, they are the gauge transformations of a Chern--Simons theory, as we show in Sec.~\ref{sec:superalgebra}.

The covariant derivative \eqref{eq:covD} transforms like $A_p$ itself. The bulk term $S_{\rm TFT}$ is therefore invariant, up to a total derivative, if each of its terms is neutral. For the weights, with $w_1=0$ for the connection itself, the kinetic terms and the cubic couplings require, respectively,
\begin{equation}
w_p+w_{10-p}=0\,,\qquad w_{2k}=w_{2k-2}+w_3\,,
\label{eq:neutral}
\end{equation}
and the parities must satisfy the same conditions modulo two.

The conditions \eqref{eq:neutral} determine the weights up to an overall constant. We fix it as in \eqref{eq:weights}, so that $w_p=\frac{5-p}{2}$ for even $p$. With this choice, the weights of the even forms can be interpreted as Weyl weights, as we discuss in Sec.~\ref{sec:bdry}. The conditions on the parities have a single nontrivial solution, which we adopt: $\eta_p=1$ for the even forms and $\eta_p=0$ for the odd ones. The $\mathrm{O}(1)$ part of \eqref{eq:Rgauge} is then
\begin{equation}
A_{2k+1}\to A_{2k+1}\,,\qquad A_{2k}\to\sigma\,A_{2k}\,,
\label{eq:O1}
\end{equation}
that is, the even forms change sign for $\sigma=-1$. In general, the weights and the parities are independent labels. For this solution, however, $2w_p$ is an integer, which is odd precisely when $\eta_p=1$. The two labels then combine, and \eqref{eq:Rgauge} takes the compact form $A_p\to\lambda^{2w_p}A_p$ for $p\neq1$, which we use below.

Since $\mathrm{O}(1)$ is discrete, it has no gauge field: the sign $\sigma$ is locally constant, and \eqref{eq:O1} does not shift $A_1$. Locally, the theories with gauge groups $\mathrm{GL}(1,\mathbb{R})$ and $\mathbb{R}^+$ are therefore the same. They differ globally: gauging $\mathrm{O}(1)$ means allowing the even forms to be glued across patches by the signs \eqref{eq:O1} (Sec.~\ref{sec:holonomy}). Algebraically, \eqref{eq:O1} is a $\mathbb{Z}_2$ grading that separates the even forms from the odd ones, and it becomes the parity automorphism of the gauge superalgebra in Sec.~\ref{sec:algebra}.

The factors of two in \eqref{eq:S11} and in \eqref{eq:Rgauge}, including the exponent $2w_p$, have a common origin, which becomes transparent if $\mathrm{GL}(1,\mathbb{R})$ is regarded as the center of $\mathrm{GL}(2,\mathbb R)$,
\begin{equation}
\lambda\mapsto g=\lambda\,\mathbf 1\,,\qquad
\det g=\lambda^2\,,\qquad
\mathrm{tr}\big(g^{-1}{\rm d} g\big)=2\,\lambda^{-1}{\rm d}\lambda\,,
\label{eq:GL2}
\end{equation}
where each two is $\mathrm{tr}\mathbf 1=2$. In terms of $g$, \eqref{eq:Rgauge} reads
\begin{equation}
A_1\to A_1+\mathrm{tr}\big(g^{-1}{\rm d} g\big)\,,\qquad A_p\to|\det g|^{w_p}\sigma^{\eta_p}A_p\quad(p\neq1)\,,
\label{eq:Rgauge2}
\end{equation}
with $\sigma=\mathrm{sgn}\,\lambda$ as in \eqref{eq:split}. Thus $A_1$ is the connection for $\det g$, and the $w_p$ are the corresponding weights. Since $\det g=\lambda^2$ is positive, it does not see the sign $\sigma$, which enters separately through the parities, as in the standard description of the characters $|\det|^{s}(\mathrm{sgn}\det)^{\eta}$ of $\mathrm{GL}(n,\mathbb R)$~\cite{Kashiwara:1978,Howe:1989,Basile:2020gqi}. Writing $\mathcal A_p=A_p\,\mathbf 1$ for $p=1,9$, the first term of \eqref{eq:S11} is a trace,
\begin{equation}
2\,\big(A_1\,{\rm d} A_9+A_9\,{\rm d} A_1\big)=\mathrm{tr}\big(\mathcal A_1\,{\rm d}\mathcal A_9+\mathcal A_9\,{\rm d}\mathcal A_1\big)\,,
\label{eq:trkin}
\end{equation}
and likewise $4A_1*A_1+A_9*A_9=\mathrm{tr}\big(2\mathcal A_1*\mathcal A_1+\frac12\mathcal A_9*\mathcal A_9\big)$ in \eqref{eq:S10}. The scalar sector of type IIB~\cite{Joung:IIB} consists of the same terms, with $A_1$ and $A_9$ valued in $\mathfrak{sl}(2,\mathbb{R})$ instead of along the identity of $\mathfrak{gl}(2,\mathbb R)=\mathfrak{sl}(2,\mathbb{R})\oplus\mathbb R\,\mathbf 1$. We do not pursue this connection here.

The bulk equations of motion are
\begin{align}
{\rm d} A_1&=0\,,\nonumber\\
{\rm D} A_3&=0\,,\nonumber\\
{\rm D} A_{2k}-A_3A_{2k-2}&=0\,,\qquad k=0,\dots,5\,,\nonumber\\
{\rm D} A_7-A_0A_8+A_2A_6-\tfrac{1}{2}\,A_4A_4&=0\,,\nonumber\\
2\,{\rm d} A_9+A_3A_7+\tfrac52A_0A_{10}-\tfrac32A_2A_8+\tfrac12A_4A_6&=0\,.
\label{eq:bulk}
\end{align}
Each line is the equation of motion of the symplectic partner of the field on which the derivative acts. These equations are of first order, contain no metric, and leave no local degrees of freedom: in a Hamiltonian treatment, their spatial components are first-class constraints generating the gauge symmetry, and the remaining components fix the evolution only up to gauge transformations. Following the usual terminology, we call them Gauss laws. The first line states that $A_1$ is flat. The last line, the equation of motion of $A_1$, is the Gauss law of the $\mathbb{R}^+$ symmetry: its source is the $\mathbb{R}^+$ current, in which each pair enters with its weight. The line with $k=5$, ${\rm D} A_{10}=A_3A_8$, which is the equation of motion of $A_0$, is an eleven-form, and therefore has no counterpart among the ten-dimensional field equations. We return to it in Secs.~\ref{sec:zeromode} and~\ref{sec:known}.

The Gauss laws \eqref{eq:bulk} are integrable: applying ${\rm D}$ to them gives identities, by the same conditions \eqref{eq:neutral} that make the action neutral. By Noether's second theorem, these identities correspond to the gauge symmetries \eqref{eq:Rgauge} and \eqref{eq:hfgauge} of $S_{\rm TFT}$, and they guarantee that these transformations map solutions of \eqref{eq:bulk} to solutions.

\subsection{Boundary term and boundary conditions}
\label{sec:bdry}

We now turn to the boundary. We first rewrite $S_{\rm TFT}$ so as to isolate its boundary part, then derive the boundary conditions implied by $S_{\rm b}$, and finally describe what happens to the gauge symmetry on the boundary.

The bulk term splits into a genuinely bulk part and an explicit boundary part. Within each symplectic pair, the two orderings of the kinetic term differ by an exact form: since $w_p+w_{10-p}=0$,
\begin{equation}
A_p{\rm D} A_{10-p}=\pm A_{10-p}{\rm D} A_p\mp{\rm d}\big(A_pA_{10-p}\big)\,,
\end{equation}
with the upper signs for odd $p$ and the lower signs for even $p$. Bringing every kinetic term to the ordering in which the higher-degree field stands in front, and collecting the cubic terms, one finds $S_{\rm TFT}=S_{\rm BF}+S_\partial$, with
\begin{align}
S_{\rm BF}&=i\int_{M_{11}}\Big[2A_9\,{\rm d} A_1+A_7\,{\rm D} A_3-A_{10}\,{\rm D} A_0+A_8\big({\rm D} A_2-A_3A_0\big)\nonumber\\
&\hspace{4em}-A_6\big({\rm D} A_4-A_3A_2\big)-\tfrac12A_3A_4A_4\Big]\,,
\label{eq:SBF}\\
S_{\partial}&=-\frac i2\int_{M_{10}}\Big[2A_1A_9+A_3A_7-A_0A_{10}+A_2A_8-A_4A_6\Big]\,.
\label{eq:Sdel}
\end{align}
We call the higher-degree fields $A_6,\dots,A_{10}$ magnetic and the lower-degree fields $A_0,\dots,A_4$ electric. In $S_{\rm BF}$, the magnetic fields enter linearly and without derivatives. They are Lagrange multipliers, whose equations of motion are the Gauss laws of the electric fields. Besides these constraints, $S_{\rm BF}$ contains only the cubic term $A_3A_4A_4$. The separation relies on the absence of a self-paired field. The boundary term $S_\partial$ accounts for the difference between $S_{\rm BF}$ and the symmetric form \eqref{eq:S11}.

The term $S_{\rm b}$ is quadratic and contains no derivatives. It is the most general such term that involves the metric: the Hodge star pairs two forms of the same degree, and there is a single field in each degree, so that only the terms $A_p*A_p$ can appear, up to coefficients. That $S_{\rm b}$ is quadratic is an assumption: since $A_0$ is a zero-form, terms of higher order such as $A_0^2\,A_p*A_p$ are compatible with $\mathrm{O}(1)$, but we do not consider them. For the even forms, they are excluded by the Weyl covariance described at the end of this section.

The boundary terms in the variation of $S_{\rm TFT}$ are
\begin{equation}
\begin{split}
\delta S_{\rm TFT}\big|_{\partial}&=\int_{M_{10}}\Big[-i\big(A_1\delta A_9+A_9\delta A_1\big)-\tfrac i2\big(A_3\delta A_7+A_7\delta A_3\big)\\
&\quad+\tfrac i2\sum_{k=0}^{5}(-1)^kA_{2k}\,\delta A_{10-2k}\Big]\,,
\end{split}
\label{eq:theta}
\end{equation}
with opposite signs for the odd and even pairs, since integrating $A_p\,{\rm d}\delta A_{10-p}$ by parts produces the boundary term $(-1)^{p}A_p\,\delta A_{10-p}$. We add the variation of $S_{\rm b}$ and leave all boundary values free. The variations with respect to the two members of a pair then give the same condition; this follows from $*^2=-1$ for the odd pairs, and from $*^2=+1$ together with the relative sign $(-1)^k$ in \eqref{eq:theta} for the even pairs. One boundary condition results for each pair,
\begin{equation}
*A_1=-\tfrac{i}{2}\,A_9\,,\qquad *A_3=-i\,A_7\,,\qquad *A_{2k}=i\,(-1)^kA_{10-2k}\,.
\label{eq:bcproj}
\end{equation}
These conditions become real in Lorentzian signature.\footnote{In Lorentzian signature, where $*^2=1$ on odd forms and $*^2=-1$ on even forms, $S_{\rm TFT}$ is real and $S_{\rm b}$ changes sign, so \eqref{eq:bcproj} holds with $-i$ replaced by $1$ (and $-\tfrac i2$ by $\tfrac12$); in particular, $*A_{2k}=(-1)^{k+1}A_{10-2k}$ (Appendix~\ref{app:conventions}).} 

The two conditions of a pair coincide only if the product of the coefficients of $A_p*A_p$ and $A_{10-p}*A_{10-p}$ in \eqref{eq:S10} equals the square of the coefficient of the pair in \eqref{eq:S11}; for $(A_1,A_9)$, this is $4\cdot1=2^2$. The coefficients of \eqref{eq:S10} satisfy this condition, which leaves one free ratio for each of the five pairs. Unlike the bulk term, which is fixed by the gauge algebra (Sec.~\ref{sec:superalgebra}), $S_{\rm b}$ is therefore not determined by the construction: the coefficients of \eqref{eq:S10} are an input, chosen so that the reduction of Sec.~\ref{sec:reduction} gives type IIA supergravity (Sec.~\ref{sec:romans}), where supersymmetry fixes them. In particular, the values $4$ and $1$ give the scalar edge mode a canonically normalized kinetic term.

Of the five free ratios, two can be removed by redefinitions. A constant $\mathbb{R}^+$ transformation \eqref{eq:Rgauge} leaves $S_{\rm TFT}$ invariant and rescales the coefficients of each pair in \eqref{eq:S10} by $\lambda^{4w_p}$ and $\lambda^{-4w_p}$, and therefore removes one combination of them. A constant rescaling $g\to\Omega^2g$ of the background metric multiplies $A_p*A_p$ by $\Omega^{10-2p}$, and removes a second, independent combination. The remaining three (four, once an Einstein--Hilbert term fixes the scale of the metric) are physical: they set the kinetic terms of the reduced action relative to its Chern--Simons terms, and the dilaton couplings relative to the kinetic term of the dilaton.

We finally turn to the gauge symmetries on the boundary. The boundary term $S_{\rm b}$ breaks $\mathrm{GL}(1,\mathbb{R})$ to $\mathrm{O}(1)\cong\mathbb{Z}_2$, even for constant $\lambda$. On a pair with $w_p\neq0$, a constant $\lambda$ acts as
\begin{equation}
(A_p,A_{10-p})\to\big(\lambda^{2w_p}A_p,\,\lambda^{-2w_p}A_{10-p}\big)\,,
\label{eq:hyperbolic}
\end{equation}
which preserves the pairing $A_pA_{10-p}$ of the bulk but rescales $A_p*A_p$ and $A_{10-p}*A_{10-p}$ by inverse factors, unless $\lambda^{4w_p}=1$. For real $\lambda$, this leaves only $\lambda=\pm1$: the $\mathbb{R}^+$ factor of \eqref{eq:split} is broken, and the $\mathrm{O}(1)$ factor is preserved. The terms of the neutral pair $(A_1,A_9)$ are invariant under constant $\lambda$, but $A_1*A_1$ is not invariant under the local transformations, which shift $A_1$. The edge modes associated with $\mathrm{GL}(1,\mathbb{R})$ therefore parametrize the coset $\mathrm{GL}(1,\mathbb{R})/\mathrm{O}(1)\cong\mathbb{R}^+$: they consist of a single scalar. Since $\mathrm{O}(1)$ is discrete, the whole of $\mathfrak{gl}(1)$ is a coset direction. This differs from type IIB~\cite{Joung:IIB}, where the unbroken $\mathrm{SO}(2)$ is continuous: its component of $A_1$ remains a connection on the boundary, and the corresponding component of $A_9$ must vanish there. Here no such condition arises, and all boundary values are left free.

The breaking of $\mathbb{R}^+$ has a different origin for the even and for the odd forms, and this difference explains the choice of weights \eqref{eq:weights}. Under a Weyl rescaling $g\to\lambda^{-2}g$ of the metric of $M_{10}$, the Hodge star acting on $p$-forms is multiplied by $\lambda^{2p-10}$. For an even form, with $2w_p=5-p$, this factor compensates exactly the factor $\lambda^{4w_p}$ that $A_p*A_p$ acquires under \eqref{eq:Rgauge}, for local $\lambda$ as well, since $S_{\rm b}$ contains no derivatives. The terms $A_{2k}*A_{2k}$ therefore break $\mathbb{R}^+$ only because the metric is a fixed background, and in this sense the weights of the even forms are their Weyl weights. The same property singles out the quadratic terms: a monomial of degree $n$ in the even forms, built with the metric and without derivatives, acquires the factor $\lambda^{5(n-2)}$ under the combined transformation, so that terms such as $A_0^2\,A_{2k}*A_{2k}$ are not invariant. The terms of the odd forms are not invariant under the combined transformation: they break $\mathbb{R}^+$ independently of the metric, and the scalar edge mode is due to them. In type IIA, these properties become the statement that the RR fields couple to the string-frame metric without the dilaton, while the NSNS sector carries the factor $e^{-2\phi}$ (Sec.~\ref{sec:dictionary}).

Under local transformations, the bulk term is not invariant either. Under the gauge transformations \eqref{eq:Rgauge} and \eqref{eq:hfgauge}, it shifts by a boundary integral,
\begin{equation}
\delta S_{\rm TFT}=\frac i2\int_{M_{10}}\Big[2A_9\,{\rm d}\varepsilon_0+2A_1\,{\rm d}\varepsilon_8+A_3\,{\rm d}\varepsilon_6+A_7\,{\rm d}\varepsilon_2-\sum_{k=0}^{4}(-1)^kA_{2k}\,{\rm d}\varepsilon_{9-2k}\Big]\,.
\label{eq:Rshift}
\end{equation}
For $\mathrm{GL}(1,\mathbb{R})$, the BF action \eqref{eq:SBF} is invariant, and the shift comes from the term $A_1A_9$ of $S_\partial$, which contains the connection itself; for a finite transformation it is $2i\int_{M_{10}}A_9\,\lambda^{-1}{\rm d}\lambda$. The shift vanishes for closed parameters, in particular for constant $\lambda$ and for $\lambda=-1\in\mathrm{O}(1)$, but not for parameters that vary along $M_{10}$. Nor do the higher-form gauge transformations preserve $S_{\rm b}$, which is not invariant under the shifts $A_p\to A_p+{\rm D}\varepsilon_{p-1}$. On $M_{10}$, the whole gauge symmetry is thus broken to $\mathrm{O}(1)$. Gauge transformations that act nontrivially on $M_{10}$ are not redundancies of the description: they change the boundary values of the fields, and the corresponding modes become physical edge modes, the boundary Goldstone modes of the broken symmetry. They will turn out to be the fields of type IIA supergravity: the edge mode on $\mathrm{GL}(1,\mathbb{R})/\mathrm{O}(1)$ is the dilaton, and the higher-form gauge parameters of the electric fields become the potentials $B_2$, $C_1$ and $C_3$ (Secs.~\ref{sec:gaugedata} and~\ref{sec:dictionary}). Their dynamics is supplied by the boundary terms $S_\partial$ and $S_{\rm b}$ (Sec.~\ref{sec:reduction}).

\section{Boundary reduction}
\label{sec:reduction}

We now reduce our action to the boundary. We start from the BF form \eqref{eq:SBF}--\eqref{eq:Sdel} of the bulk term, in which the magnetic fields are Lagrange multipliers. We eliminate them, in the bulk and on the boundary (Sec.~\ref{sec:eliminate}), and then solve the constraints that they impose (Sec.~\ref{sec:solve}). The integration data of the solution are the fields of the boundary theory, and we relate them to the gauge symmetries of the bulk (Sec.~\ref{sec:gaugedata}).

\subsection{Eliminating the magnetic fields}
\label{sec:eliminate}

The magnetic fields enter $S_{\rm bTFT}=S_{\rm BF}+S_\partial+S_{\rm b}$ algebraically: linearly in the bulk, through \eqref{eq:SBF}, and on the boundary linearly through \eqref{eq:Sdel} and quadratically through $S_{\rm b}$. They can therefore be eliminated exactly by their own equations of motion. In the bulk, these are the Gauss laws of the electric fields,
\begin{equation}
{\rm d} A_1=0\,,\qquad {\rm D} A_3=0\,,\qquad {\rm D} A_0=0\,,\qquad
{\rm D} A_2=A_3A_0\,,\qquad {\rm D} A_4=A_3A_2\,,
\label{eq:constraints}
\end{equation}
on which $S_{\rm BF}$ reduces to its last term. On the boundary, they are the boundary conditions \eqref{eq:bcproj}, which we solve for the magnetic fields,
\begin{equation}
A_9=2i*A_1\,,\qquad A_7=i*A_3\,,\qquad A_{10-2k}=-i(-1)^k*A_{2k}\,.
\label{eq:magnetic}
\end{equation}
Substituting \eqref{eq:magnetic} into \eqref{eq:Sdel} turns every term of $S_\partial$ into a positive kinetic term,
\begin{equation}
S_\partial\big|_{\eqref{eq:magnetic}}=\int_{M_{10}}\Big[2A_1*A_1+\tfrac12A_3*A_3+\tfrac12\sum_{k=0}^{2}A_{2k}*A_{2k}\Big]\,,
\end{equation}
while $S_{\rm b}$ vanishes: on \eqref{eq:magnetic}, $A_9*A_9=-4A_1*A_1$ cancels $4A_1*A_1$, and likewise within each of the other pairs. The total action therefore becomes
\begin{equation}
S_{\rm bTFT}=\int_{M_{10}}\Big[2\,A_1*A_1+\tfrac{1}{2}\,A_3*A_3+\tfrac{1}{2}\sum_{k=0}^{2}A_{2k}*A_{2k}\Big]-\tfrac{i}{2}\int_{M_{11}}A_3A_4A_4\,,
\label{eq:Sred}
\end{equation}
with the electric fields subject to \eqref{eq:constraints}. All kinetic terms come from the boundary term $S_\partial$ of $S_{\rm TFT}$, and the role of $S_{\rm b}$ is only to select the boundary conditions \eqref{eq:bcproj}. The metric nevertheless enters \eqref{eq:Sred} in the standard way. Since $S_{\rm b}$ is the only metric-dependent term of $S_{\rm bTFT}$, and $S_{\rm bTFT}$ is stationary with respect to the magnetic fields on \eqref{eq:magnetic}, the variation of $S_{\rm b}$ with respect to the metric, evaluated on \eqref{eq:bcproj}, equals that of \eqref{eq:Sred}: the energy--momentum tensor of our action is that of the reduced action.

Equation~\eqref{eq:Sred} is the central result of the reduction. The magnetic fields have been eliminated exactly, through their own equations of motion, and no equation has been imposed on the electric fields or on the dilaton. Equation~\eqref{eq:Sred} is therefore an off-shell action for the electric fields alone, subject to the constraints \eqref{eq:constraints}. Its kinetic terms live on $M_{10}$, and its only bulk term is the cubic term, which becomes the Chern--Simons terms on $M_{10}$ (Sec.~\ref{sec:fromboundary}). In this sense, the dilaton and form-field sector of massive type IIA is entirely an edge mode of the topological theory on $M_{11}$. That type IIB string theory could be an edge mode of a topological theory in higher dimensions, with a bulk in $10+2$ dimensions, was conjectured in Ref.~\cite{Heckman:2017uxe}; we compare with the eleven-dimensional bulk of type IIB~\cite{Joung:IIB} in Sec.~\ref{sec:discussion}.

\subsection{Solving the constraints}
\label{sec:solve}

From the point of view of the bulk theory, the constraints \eqref{eq:constraints} are differential equations for the electric fields. We solve them locally, in two steps. First, the flatness of $A_1$ is solved by a function ${\cal V}$ with values in $\mathbb{R}^+$, the identity component of the gauge group $\mathrm{GL}(1,\mathbb{R})$, just as the flatness of the $\mathrm{SL}(2,\mathbb{R})$ connection is solved by an $\mathrm{SL}(2,\mathbb{R})$-valued function in type IIB~\cite{Joung:IIB}. We then write each of the other electric fields as a power of ${\cal V}$, fixed by its weight, times a new form,
\begin{equation}
A_1={\cal V}^{-1}{\rm d}{\cal V}\,,\qquad A_3={\cal V}^{-1}H_3\,,\qquad A_{2k}={\cal V}^{\frac{5-2k}{2}}F_{2k}\,.
\label{eq:puregauge}
\end{equation}
Since ${\rm D}\big({\cal V}^{w}F\big)={\cal V}^{w}{\rm d} F$ for a form of weight $w$, the constraints \eqref{eq:constraints} become equations for the new forms alone,
\begin{equation}
{\rm d} H_3=0\,,\qquad {\rm d} F_0=0\,,\qquad {\rm d} F_2=H_3F_0\,,\qquad {\rm d} F_4=H_3F_2\,.
\label{eq:HFconstraints}
\end{equation}
Second, these equations are solved locally by integration, which introduces a two-form $B_2$, a one-form $C_1$, a three-form $C_3$ and a constant $m$,
\begin{equation}
H_3={\rm d} B_2\,,\qquad F_0=m\,,\qquad F_2={\rm d} C_1+m\,B_2\,,\qquad F_4={\rm d} C_3-H_3\,C_1+\tfrac m2\,B_2^2\,.
\label{eq:HFsol}
\end{equation}
In terms of the polyforms $F=F_0+F_2+F_4$ and $C=C_1+C_3$, the solution reads
\begin{equation}
F={\rm d}_HC+m\,e^{B_2}\,,\qquad {\rm d}_H={\rm d}-H_3\,,
\label{eq:polysol}
\end{equation}
where only the components of degree up to four are kept on the right-hand side; it solves ${\rm d}_HF=0$, the polyform version of \eqref{eq:HFconstraints}. The constant $m$ arises from ${\rm d} F_0=0$, that is, from ${\rm D} A_0=0$, the equation of motion of $A_{10}$.

The reduced action \eqref{eq:Sred} is thus a functional of ${\cal V}$, of $B_2$, $C_1$ and $C_3$, and of $m$. The notation anticipates the interpretation of Sec.~\ref{sec:fromboundary}, where $H_3$ and $F_{2k}$ become the field strengths of massive type IIA. Here, \eqref{eq:HFconstraints} are simply the Gauss laws of the electric fields, and \eqref{eq:HFsol} is their local solution.

\subsection{The integration data as gauge parameters}
\label{sec:gaugedata}

The solution \eqref{eq:puregauge}, \eqref{eq:HFsol} is not unique, and the ambiguity has a simple origin. Apart from $m$, the integration data are parameters of gauge transformations. They can be removed locally in the bulk, and become physical only on the boundary, where the gauge symmetry is broken (Sec.~\ref{sec:bdry}). The constant $m$ is not of this kind, since a zero-form has no gauge parameter. It is the covariantly constant mode of $A_0$, and a variable of the theory like the other fields; we derive its equation of motion in Sec.~\ref{sec:zeromode}.

For ${\cal V}$, the gauge transformation is the $\mathbb{R}^+$ part of \eqref{eq:Rgauge}, which acts as ${\cal V}\to\lambda^2\,{\cal V}$ at fixed $H_3$ and $F_{2k}$. The $\mathrm{O}(1)$ part does not act on ${\cal V}$; it changes the sign of the forms $F_{2k}$. In the Weyl language of Sec.~\ref{sec:bdry}, the connection $A_1$ plays the role of the dilatation gauge field of conformal gravity~\cite{Kaku:1978nz}. Here, it is abelian and flat, that is, Weyl-integrable~\cite{Dirac:1973gk}, and ${\cal V}$ is the St\"uckelberg mode of the dilatation. The Liouville field arises in the same way from the dilatation component of the $\mathfrak{so}(2,2)$ connection of three-dimensional gravity~\cite{Joung:2025edge}. There, the dilatation connection is not flat, and the Liouville mode is extracted with a Dirichlet condition compensated by the special conformal component. Here, no such component exists, and ${\cal V}$ acquires its kinetic term $2A_1*A_1$ in \eqref{eq:Sred} from the boundary term $S_\partial$.

For $B_2$, $C_1$ and $C_3$, the gauge transformations are those of the electric fields that preserve the Gauss laws \eqref{eq:constraints},
\begin{equation}
\delta A_3={\rm D}\varepsilon_2\,,\qquad
\delta A_2={\rm D}\varepsilon_1+\varepsilon_2A_0\,,\qquad
\delta A_4={\rm D}\varepsilon_3+\varepsilon_2A_2-A_3\,\varepsilon_1\,,
\label{eq:elgauge}
\end{equation}
where the parameter $\varepsilon_{p-1}$ is a $(p-1)$-form with the weight of $A_p$. They are the electric part of \eqref{eq:hfgauge}. Writing $\varepsilon_{p-1}={\cal V}^{w_p}\lambda_{p-1}$, they act on the solution as shifts of the integration data,
\begin{equation}
\delta B_2=\lambda_2\,,\qquad \delta C_1=\lambda_1\,,\qquad \delta C_3=\lambda_3+\lambda_2\,C_1\,.
\label{eq:shift}
\end{equation}

The parameters of \eqref{eq:elgauge} are themselves defined only up to
\begin{equation}
\varepsilon_2\to\varepsilon_2+{\rm D}\xi_1\,,\qquad
\varepsilon_1\to\varepsilon_1+{\rm D}\xi_0-\xi_1A_0\,,\qquad
\varepsilon_3\to\varepsilon_3+{\rm D}\xi_2-\xi_1A_2-A_3\,\xi_0\,,
\label{eq:gfg}
\end{equation}
where $\xi_{p-2}$ is a $(p-2)$-form with the weight of $A_p$. By the Gauss laws \eqref{eq:constraints}, this change of the parameters does not affect \eqref{eq:elgauge}. Such gauge-for-gauge transformations do not act on the fields $A_p$, but through \eqref{eq:shift} they act on the integration data. We return to them in Sec.~\ref{sec:fromboundary}.

\section{Massive type IIA and the Romans mass}
\label{sec:romans}

We now show that the result of Sec.~\ref{sec:reduction} is massive type IIA supergravity. We first evaluate the reduced action and find Romans' action (Sec.~\ref{sec:fromboundary}). We then keep the magnetic fields and recover the democratic formulation (Sec.~\ref{sec:democratic}). Next, we identify the symmetries of our action in type IIA (Sec.~\ref{sec:dictionary}). Finally, we show that the Romans mass is varied, and derive its equation of motion (Sec.~\ref{sec:zeromode}).

\subsection{The reduced action is Romans' action}
\label{sec:fromboundary}

The reduced action \eqref{eq:Sred} is the sum of a boundary integral and a bulk integral. We evaluate them in turn on the solution \eqref{eq:puregauge}, \eqref{eq:HFsol}, and parametrize the positive function ${\cal V}$ as ${\cal V}=e^{\phi/2}$.

For the boundary integral, \eqref{eq:puregauge} becomes
\begin{equation}
A_1=\tfrac12\,{\rm d}\phi\,,\qquad A_3=e^{-\frac12\phi}H_3\,,\qquad A_{2k}=e^{\frac{5-2k}{4}\phi}F_{2k}\,,
\label{eq:dressed}
\end{equation}
so that
\begin{align}
&\int_{M_{10}}\Big[2\,A_1*A_1+\tfrac{1}{2}\,A_3*A_3+\tfrac{1}{2}\sum_{k=0}^{2}A_{2k}*A_{2k}\Big]\nonumber\\
&\qquad=\int_{M_{10}}\Big[\tfrac12\,{\rm d}\phi*{\rm d}\phi+\tfrac12\,e^{-\phi}H_3*H_3+\tfrac12\sum_{k=0}^{2}e^{\frac{5-2k}{2}\phi}F_{2k}*F_{2k}\Big]\,.
\label{eq:kinred}
\end{align}
The function $\phi$ acquires a canonical kinetic term, and each weight becomes an exponent: a field of weight $w_p$ couples to $\phi$ through $e^{w_p\phi}$.

For the bulk integral, the cubic term is neutral, so that ${\cal V}$ drops out of it: $A_3A_4A_4=H_3F_4F_4$. On \eqref{eq:HFsol}, this eleven-form is exact,
\begin{equation}
H_3F_4F_4={\rm d}\Big[B_2\,{\rm d} C_3\,{\rm d} C_3+\tfrac m3\,B_2^3\,{\rm d} C_3+\tfrac{m^2}{20}\,B_2^5\Big]\,.
\label{eq:descent}
\end{equation}
Indeed, $H_3H_3=0$ removes every term that contains $C_1$, and ${\rm d}(B_2^n)=nH_3B_2^{n-1}$ for the even form $B_2$. By Stokes' theorem, the bulk integral therefore reduces to an integral over $M_{10}$.

Adding the two contributions, the reduced action is
\begin{align}
S_{\rm bTFT}\big|_{\eqref{eq:HFsol}}&=\int_{M_{10}}\Big[\tfrac{1}{2}\,{\rm d}\phi*{\rm d}\phi+\tfrac{1}{2}\,e^{-\phi}H_3*H_3+\tfrac{1}{2}\sum_{k=0}^{2}e^{\frac{5-2k}{2}\phi}F_{2k}*F_{2k}\Big]\nonumber\\
&\quad-\tfrac{i}{2}\int_{M_{10}}\Big[B_2\,{\rm d} C_3\,{\rm d} C_3+\tfrac{m}{3}\,B_2^3\,{\rm d} C_3+\tfrac{m^2}{20}\,B_2^5\Big],
\label{eq:SR}
\end{align}
with the field strengths \eqref{eq:HFsol}. This is the dilaton and form-field sector of the bosonic action of massive type IIA supergravity~\cite{Romans:1985tz} in the Einstein frame, that is, Romans' action without the Einstein--Hilbert term, the metric being a background here. The function $\phi$ is the dilaton, the forms $B_2$, $C_1$ and $C_3$ are the gauge potentials, and the constant $m$ is the Romans mass, which is here a variable of the action (Sec.~\ref{sec:zeromode}).

The reduction thus reproduces this sector of Romans' action off shell, and each of its terms has a definite origin in our action. The kinetic terms come from the boundary term $S_\partial$ of $S_{\rm TFT}$, with the dilaton couplings fixed by the weights. Their ratios, $\frac52:\frac32:\frac12:-1$ for $F_0$, $F_2$, $F_4$ and $H_3$, follow from the conditions \eqref{eq:neutral} and do not depend on the coefficients of $S_{\rm b}$. The $k=0$ term of the sum is the Romans potential $\frac12 m^2e^{5\phi/2}$. The St\"uckelberg couplings $m\,B_2$ and $\frac m2B_2^2$ of the field strengths \eqref{eq:HFsol} come from the sources $A_3A_{2k-2}$ in the Gauss laws \eqref{eq:constraints}. The Chern--Simons terms, with the relative coefficients $1:\frac m3:\frac{m^2}{20}$, come from the single term of $S_{\rm BF}$ that contains no multiplier. That they descend from the eleven-form $H_3F_4F_4$ was observed in Ref.~\cite{Bergshoeff:1995super8}; here, this eleven-form is a term of the bulk action.

The gauge symmetries of this action also originate in the bulk. The gauge transformations \eqref{eq:elgauge} are broken on the boundary, where they shift the potentials. The gauge-for-gauge transformations \eqref{eq:gfg}, in contrast, do not act on the fields $A_p$, and therefore remain symmetries of the reduced action \eqref{eq:Sred}. For $\xi_1={\cal V}^{-1}\Lambda_1$ and $\xi_0=\xi_2=0$, they act on the potentials through \eqref{eq:shift} as
\begin{equation}
\delta B_2={\rm d}\Lambda_1\,,\qquad \delta C_1=-m\,\Lambda_1\,,\qquad \delta C_3=-m\,B_2\,\Lambda_1\,,
\label{eq:Stueck}
\end{equation}
up to an exact term ${\rm d}(\Lambda_1C_1)$ in $\delta C_3$. The field strengths \eqref{eq:HFsol} are invariant, and the integrand of the Chern--Simons terms changes by the exact form ${\rm d}\big[\Lambda_1\,{\rm d} C_3\,{\rm d} C_3-m\,B_2^2\,\Lambda_1\,{\rm d} C_3-\tfrac{m^2}{12}\,B_2^4\,\Lambda_1\big]$. This is the St\"uckelberg symmetry of massive type IIA, by which the two-form absorbs the one-form and becomes massive. The shift of $C_1$ by $-m\,\Lambda_1$ is the term $-\xi_1A_0$ of \eqref{eq:gfg}, which reflects the source $A_3A_0$ in the Gauss law of $A_2$. The ordinary gauge transformations of the RR potentials, $\delta C_1={\rm d}\Lambda_0$ with $\delta C_3=-H_3\Lambda_0$, and $\delta C_3={\rm d}\Lambda_2$, arise in the same way, from $\xi_0={\cal V}^{3/2}\Lambda_0$ and $\xi_2={\cal V}^{1/2}\Lambda_2$.

The $B_2^5$ term deserves a comment, since it exists only because the reduction keeps the electric fields. In $S_{\rm TFT}$, the field $A_3$ multiplies the full source of $A_7$ in \eqref{eq:bulk},
\begin{equation}
A_0A_8-A_2A_6+\tfrac12\,A_4A_4\,,
\label{eq:A7source}
\end{equation}
and not only its last term. If the Gauss laws of $A_6$ and $A_8$ were solved in the same way as those of the electric fields, so that \eqref{eq:polysol} held in all degrees, the terms without $C$, $A_{2k}={\cal V}^{w_{2k}}\,m\,B_2^k/k!$, would cancel in \eqref{eq:A7source},
\begin{equation}
m^2B_2^4\,\Big(\frac1{0!\,4!}-\frac1{1!\,3!}+\frac1{2}\,\frac1{2!\,2!}\Big)=0\,,
\label{eq:B4cancel}
\end{equation}
since the Mukai pairing of $e^{B_2}$ with itself vanishes (Sec.~\ref{sec:superalgebra}), and no $B_2^5$ term would arise. In the reduction of Sec.~\ref{sec:reduction}, however, $A_6$ and $A_8$ are Lagrange multipliers, and the terms that contain them vanish on the constraints \eqref{eq:constraints}. Only the last term of \eqref{eq:A7source} remains in \eqref{eq:Sred}, and the part $\frac m2B_2^2$ of $F_4$ contributes $H_3\,\big(\tfrac m2\,B_2^2\big)^2={\rm d}\big[\tfrac{m^2}{20}\,B_2^5\big]$ to \eqref{eq:descent}.

More generally, our action itself contains no Chern--Simons terms. They are the form that the invariant cubic term takes on the boundary once the electric fields are kept and the magnetic ones eliminated. This is the counterpart, in our action, of the absence of Chern--Simons terms in the democratic pseudo-action of type II supergravity~\cite{Bergshoeff:2001pv}, where they appear only once the dual field strengths are eliminated. The bulk form is also the more general one. The descent \eqref{eq:descent} uses the potentials $B_2$ and $C_3$, and is therefore local. When $H_3$ or $F_4$ has a nontrivial period, the potentials exist only in patches, and the integrand of the Chern--Simons terms of \eqref{eq:SR} is not a globally defined ten-form. The bulk integral of $A_3A_4A_4$ in \eqref{eq:Sred}, which involves only the field strengths, then provides their definition, as for any Chern--Simons term defined through a bounding manifold. This requires the field strengths to extend from $M_{10}$ over $M_{11}$, and the result may depend on the extension; this dependence belongs to the quantum definition of the theory (Sec.~\ref{sec:discussion}).

\subsection{The democratic formulation}
\label{sec:democratic}

The reduction eliminated the magnetic fields. We now keep them, and show that the full set of equations of our action, the Gauss laws \eqref{eq:bulk} and the boundary conditions \eqref{eq:bcproj}, is the democratic formulation of massive type IIA.

When all the Gauss laws are imposed, and not only those of the electric fields, the solution \eqref{eq:puregauge} extends to all fields,
\begin{equation}
A_1={\cal V}^{-1}{\rm d}{\cal V}\,,\qquad A_p={\cal V}^{w_p}F_p\,,\qquad A_9=F_9\,,
\label{eq:dict}
\end{equation}
for $p\neq1,9$, with $F_3=H_3$ and $F_7=H_7$. Each term of the Gauss laws \eqref{eq:bulk} is neutral, so that ${\cal V}$ drops out. The first line, ${\rm d} A_1=0$, holds identically on \eqref{eq:dict}, and the others become, line by line,
\begin{align}
{\rm d} H_3&=0,\nonumber\\
{\rm d} F_{2k}-H_3\,F_{2k-2}&=0,\qquad k=0,\dots,5,\nonumber\\
{\rm d} H_7-F_0F_8+F_2F_6-\tfrac{1}{2}\,F_4F_4&=0,\nonumber\\
2\,{\rm d} F_9+H_3H_7+\tfrac{5}{2}\,F_0F_{10}-\tfrac{3}{2}\,F_2F_8+\tfrac{1}{2}\,F_4F_6&=0,
\label{eq:bianchi}
\end{align}
with $F_{-2}\equiv0$. Similarly, with ${\cal V}=e^{\phi/2}$, the boundary conditions \eqref{eq:bcproj} become
\begin{equation}
*{\rm d}\phi=-i\,F_9\,,\qquad e^{-\phi}*H_3=-i\,H_7\,,\qquad
e^{\frac{5-2k}{2}\phi}*F_{2k}=i\,(-1)^kF_{10-2k}\,.
\label{eq:dual}
\end{equation}
For instance, the condition $*A_3=-i\,A_7$ reads $e^{-\frac12\phi}*H_3=-i\,e^{\frac12\phi}H_7$.

These are the equations of democratic type IIA supergravity~\cite{Bergshoeff:2001pv}. There, \eqref{eq:dual} are the duality relations, which define the dual field strengths $F_6$, $F_8$, $F_{10}$, $H_7$ and $F_9$, and they turn the field equations of \eqref{eq:SR} into the Bianchi identities \eqref{eq:bianchi}. In the second line of \eqref{eq:bianchi}, the cases $k\le2$ are the identities \eqref{eq:HFconstraints} solved by \eqref{eq:HFsol}, the cases $k=3,4$ are the equations of motion of $C_3$ and $C_1$, and the case $k=5$ is an eleven-form, whose content is global (Sec.~\ref{sec:zeromode}). The third and the last lines are the equations of motion of $B_2$ and $\phi$. The boundary values of the bulk fields are thus the dilaton-dressed field strengths of democratic massive type IIA, and the symplectic pairs $(A_p,A_{10-p})$ are its duality pairs~\cite{Bergshoeff:2001pv,Bergshoeff:2006qw}. The correspondence is summarized in Table~\ref{tab:dict}.

\begin{table}[t]\centering
\caption{Our action and massive type IIA supergravity.}
\label{tab:dict}
\small
\begin{tabular}{@{}ll@{}}
\toprule
our action & massive type IIA\\
\midrule
$A_0$, $A_2$, $A_4$, $A_6$, $A_8$, $A_{10}$ & RR field strengths $F_0,\dots,F_{10}$, dressed by ${\cal V}^{w_p}$\\
$A_3$, $A_7$ & NSNS field strengths $H_3$, $H_7$, dressed by ${\cal V}^{w_p}$\\
$A_1$, $A_9$ & $\frac12{\rm d}\phi$ and the dual $F_9$\\
weights $w_p$ & dilaton couplings, $A_p*A_p=e^{w_p\phi}F_p*F_p$\\
symplectic pairs $(A_p,A_{10-p})$ & duality pairs; $(A_0,A_{10})$: Romans mass and $F_{10}$\\
Gauss laws \eqref{eq:bulk} & Bianchi identities \eqref{eq:bianchi}\\
\quad electric, solved in Sec.~\ref{sec:solve} & \quad identities, solved by $B_2$, $C_1$, $C_3$ and $m$\\
\quad magnetic & \quad field equations of \eqref{eq:SR}\\
\quad Gauss law of $\mathbb{R}^+$ & \quad dilaton equation; $\frac52A_0A_{10}$: Romans potential\\
\quad Gauss law of degree eleven & \quad equation of motion of $m$, \eqref{eq:dSdm}\\
boundary conditions \eqref{eq:bcproj} & duality relations \eqref{eq:dual}\\
cubic coupling $A_3A_{2k}A_{8-2k}$ & St\"uckelberg couplings and Chern--Simons terms\\
$\mathbb{R}^+\subset\mathrm{GL}(1,\mathbb{R})$, broken by $S_{\rm b}$ & rigid shift of the dilaton at fixed Einstein metric\\
$\mathrm{O}(1)\subset\mathrm{GL}(1,\mathbb{R})$, preserved by $S_{\rm b}$ & $(-1)^{F_L}$\\
edge modes on $\mathrm{GL}(1,\mathbb{R})/\mathrm{O}(1)$ & dilaton\\
\bottomrule
\end{tabular}
\end{table}

The two sides differ in the status of two equations. The flatness of $A_1$ and the constancy of $A_0$ are Gauss laws of our action, the equations of motion of $A_9$ and $A_{10}$. In supergravity they are identities, since $\phi$ is a scalar and $m$ a parameter: the value of $m$ must be fixed before the action \eqref{eq:SR} is written, and a D8-brane, across which $m$ jumps, separates regions described by different actions. The formulations with a nine-form potential~\cite{Bergshoeff:1995super8,Bergshoeff:2001pv,Bandos:2003et} avoid this. The reduction of Sec.~\ref{sec:reduction} keeps $A_0$, so that $m$ enters the reduced action as a constant. Keeping $A_{10}$ instead, whose Gauss law ${\rm D} A_{10}=A_3A_8$ is solved by a nine-form potential, should lead to these formulations, in which the mass is a field and D8-branes are electric sources. The eleven-form Gauss law would then be solved explicitly, and the condition \eqref{eq:page} would hold identically, since $(e^{-B_2}F)_{10}$ would be exact. That a mass parameter and a top-form field strength are canonically conjugate is well known~\cite{Henneaux:1984ji,Bergshoeff:1995super8}; here, the two descriptions would be two reductions of the same bulk (Sec.~\ref{sec:twist}). With $A_0$ kept, there is a further difference, of a global nature: $m$, although constant, is varied. We derive its equation of motion in Sec.~\ref{sec:zeromode}, after comparing the symmetries.

Other reductions are possible as well. As for type IIB~\cite{Joung:IIB}, one may single out a time direction and reduce in Hamiltonian form, which leads to a duality-symmetric action that is not manifestly Lorentz invariant, of the type of Refs.~\cite{Henneaux:1988gg,Schwarz:1993vs}. A covariant reduction, along the lines of Ref.~\cite{Arvanitakis:2022bnr}, leads instead to a democratic action with auxiliary fields, of the form of Ref.~\cite{Mkrtchyan:2022xrm}, and eliminating the auxiliary fields gives an action of PST type, which for massive type IIA was constructed in Ref.~\cite{Bandos:2003et}; its relation to the superalgebra of the doubled formalism was discussed, in the massless case, in Ref.~\cite{Nurmagambetov:2004}. Neither these reductions nor the one keeping $A_{10}$ is carried out here.

In type IIA string theory, the Bianchi identity of $H_7$, the third line of \eqref{eq:bianchi}, receives a further term $X_8=\frac{1}{192}\big(p_1^2-4\,p_2\big)$, built from the Pontryagin forms of the metric. It comes from the one-loop coupling $B_2\wedge X_8$~\cite{Vafa:1995fj,Duff:1995wd}, which cancels by inflow the anomaly of the NS5-brane world-volume. In the eleven-dimensional action of Ref.~\cite{Apruzzi:2023uma}, where no dilaton scaling is gauged, it is added as a term $H_3\wedge X_8$, with $X_8$ a prescribed eight-form. In our action, the corresponding term $A_3\wedge X_8$ cannot simply be added: since $X_8$ carries no weight, the term is not neutral, and it would break the $\mathbb{R}^+$ gauge symmetry and with it the integrability of the Gauss laws. This is consistent with the origin of the coupling: the $\mathbb{R}^+$ scaling is a symmetry of the classical two-derivative theory, and the one-loop term breaks it, as does the quantization of the fluxes (Sec.~\ref{sec:discussion}). The $\mathrm{O}(1)$ symmetry is preserved, since $A_3$ has no parity. Including the term would therefore require a mechanism that breaks $\mathbb{R}^+$ to $\mathrm{O}(1)$, as well as the metric in the bulk, since $X_8$ is built from the curvature, whereas here the metric is a fixed background that enters only through the boundary term. We leave this, together with the world-volume theory of the NS5-brane, for future work.

\subsection{Symmetries}
\label{sec:dictionary}

We now compare the symmetries. As anticipated in Sec.~\ref{sec:bdry}, the gauge directions broken on the boundary have become fields of type IIA: the edge mode on $\mathrm{GL}(1,\mathbb{R})/\mathrm{O}(1)$ is the dilaton, and the higher-form gauge parameters of the electric fields are the potentials $B_2$, $C_1$ and $C_3$ (Sec.~\ref{sec:gaugedata}). What remains to compare is the rigid part. In our action, $\mathrm{GL}(1,\mathbb{R})\cong\mathbb{R}^+\times\mathrm{O}(1)$ is a gauge symmetry, which the boundary term breaks to $\mathrm{O}(1)$. In type IIA, the same group acts by rigid transformations. Type IIA supergravity has two rigid scaling symmetries~\cite{Bergshoeff:2002mb}: one leaves the action invariant and shifts the dilaton at fixed Einstein metric, and the other rescales both the metric and the action. Since the Einstein metric is a background here, only the former is relevant. Together with a sign, it acts, for a constant $\lambda_0\in\mathrm{GL}(1,\mathbb{R})$, as
\begin{equation}
{\cal V}\to\lambda_0^2\,{\cal V}\,,\qquad F_p\to\lambda_0^{-2w_p}F_p\,.
\label{eq:rigid}
\end{equation}
We discuss the two factors of $\mathrm{GL}(1,\mathbb{R})$ in turn.

For $\lambda_0>0$, \eqref{eq:rigid} is the shift $\phi\to\phi+4\log\lambda_0$ of the dilaton, accompanied by a rescaling of the field strengths. In the massive theory, this rescaling includes the mass, $m\to\lambda_0^{-5}\,m$, so that \eqref{eq:rigid} relates theories with different values of $m$. The transformation leaves every $A_p={\cal V}^{w_p}F_p$ invariant: in our action, it does not act on the fields at all. The $\mathbb{R}^+$ gauge transformations \eqref{eq:Rgauge} are different. They act on ${\cal V}$ at fixed $F_p$, and they are broken on the boundary, where ${\cal V}=e^{\phi/2}$ becomes the dilaton, the edge mode on $\mathrm{GL}(1,\mathbb{R})/\mathrm{O}(1)\cong\mathbb{R}^+$. Since ${\cal V}$ enters only through the solution \eqref{eq:puregauge}, and neither $S_{\rm bTFT}$ nor its equations of motion refer to it, our action does not distinguish type IIA configurations related by \eqref{eq:rigid}, as in type IIB~\cite{Joung:IIB}. Classically, the asymptotic value of the dilaton is therefore not a parameter of our action, and only the invariant combination $A_0=e^{5\phi/4}m$ is.

The Weyl interpretation of Sec.~\ref{sec:bdry} becomes transparent in the string frame. Let $g$ be the Einstein-frame metric used so far, and $g_S={\cal V}\,g=e^{\phi/2}g$ the string-frame metric. On $p$-forms, the two Hodge stars are related by $*_S={\cal V}^{5-p}*$, so that
\begin{equation}
A_{2k}*A_{2k}=F_{2k}*_SF_{2k}\,,\qquad A_3*A_3=e^{-2\phi}\,H_3*_SH_3\,.
\label{eq:stringframe}
\end{equation}
The RR fields thus depend on ${\cal V}$ and $g$ only through $g_S$: in the string frame, their kinetic terms carry no dilaton. This is the meaning of their Weyl weights, since $g_S$ is invariant under the gauge transformation ${\cal V}\to\lambda^2\,{\cal V}$ combined with the Weyl rescaling $g\to\lambda^{-2}g$. The NSNS field strength carries in addition the factor $e^{-2\phi}={\cal V}^{-4}$, which is why the weights of $A_3$ and $A_7$ differ from Weyl weights by $\mp2$. Conversely, the Einstein-frame metric $g={\cal V}^{-1}g_S$ is invariant under the rigid transformation \eqref{eq:rigid}, under which $g_S\to\lambda_0^2\,g_S$, which is why it can be held fixed.

For $\lambda_0=-1$, \eqref{eq:rigid} leaves $\phi$ and $H_3$ invariant and changes the sign of the even forms $F_{2k}$, which are the RR field strengths. In our action, this is the $\mathrm{O}(1)$ transformation \eqref{eq:O1}, which the boundary term preserves. In type IIA, it is the action of $(-1)^{F_L}$ on the bosonic fields~\cite{Sen:1996na,Dabholkar:1997zd}. In the type II string, $F_L$ is the spacetime fermion number of the left-moving sector of the worldsheet, so that $(-1)^{F_L}$ acts as $-1$ on the states whose left-moving part is in the Ramond sector: the NSNS fields are even and the RR fields are odd. It is a symmetry of perturbative type II string theory and of the supergravity actions, in which the RR fields enter in pairs. In our action, this is the condition on the parities of Sec.~\ref{sec:weights}, that every term of $S_{\rm bTFT}$ contains an even number of even forms. The $\mathrm{O}(1)$ factor of $\mathrm{GL}(1,\mathbb{R})$ is thus generated by $(-1)^{F_L}$. On the bosonic fields, $(-1)^{F_L}$ and $(-1)^{F_R}$ act in the same way, since they differ by $(-1)^F$; without fermions, the label $F_L$ is a convention. In M-theory, this element is the reflection of the eleventh circle accompanied by $C_3\to-C_3$, under which $C_3$ and the Kaluza--Klein vector $C_1$ are odd while $B_2$ and the dilaton are even~\cite{Gaberdiel:2001xm,Aharony:2007du}.

Three properties of $(-1)^{F_L}$ will be used below. First, it reverses all RR charges, and thus exchanges D-branes with anti-D-branes. Second, it reverses the Romans mass, $m\to-m$. In Romans' supergravity, where $m$ is a parameter, it is therefore not a symmetry at fixed $m\neq0$, but maps the theory with mass $m$ to the theory with mass $-m$. In our action, $m$ is instead the integration constant of $A_0$, which is varied (Sec.~\ref{sec:zeromode}), so that both signs of $m$ occur in a single action, and $(-1)^{F_L}$ is a symmetry of it: the constant transformation $\lambda=-1$ of the residual $\mathrm{O}(1)$ gauge symmetry, which maps a configuration with mass $m$ to one with mass $-m$ and all RR fields reversed. Third, beyond this constant transformation, gauging $\mathrm{O}(1)$ allows the fields on $M_{10}$ to be glued by $(-1)^{F_L}$ across patches (Sec.~\ref{sec:holonomy}); this differs from the worldsheet orbifold by $(-1)^{F_L}$, which maps type IIA to type IIB string theory~\cite{Sen:1996na,Dabholkar:1997zd}.

The action has one further discrete symmetry when the background admits a reflection. Let $I$ be an involution of $M_{11}$ that reverses the orientation and restricts to an isometry of the metric on $M_{10}$, such as $I_9:x^9\to-x^9$. Then
\begin{equation}
\tau:\ A_p\mapsto s_p\,I^*A_p\,,\qquad s_3=s_9=-1\,,\quad s_{2k}=(-1)^k\,,\quad s_1=s_7=1\,,
\label{eq:orrev}
\end{equation}
is a symmetry of $S_{\rm bTFT}$. The signs $s_p$ change the sign of every term of $S_{\rm TFT}$, as one checks term by term, and leave $S_{\rm b}$ invariant; the pull-back by $I$ does the same, since $S_{\rm TFT}$ is the integral of a differential form, whereas $S_{\rm b}$ is built with the Hodge star. Neither is a symmetry by itself, but their combination is, and it preserves the boundary conditions \eqref{eq:bcproj} for the same reason. Unlike $\mathrm O(1)$, which is a gauge symmetry on any $M_{11}$, $\tau$ is a symmetry only of backgrounds that admit such an $I$, as parity is in a field theory.

Since the signs $s_p$ commute with $\mathrm{GL}(1,\mathbb R)$, $\tau$, which generates an orientation-reversing $\mathbb Z_2$, and $\mathrm O(1)=\{+1,-1\}$ generate a group $\mathbb Z_2\times\mathbb Z_2=\{\pm1,\pm\tau\}$. Only its orientation-preserving subgroup $\mathrm O(1)$ is gauged in our action. In type IIA string theory, where $-1\in\mathrm O(1)$ is $(-1)^{F_L}$, and with $I=I_9$, the symmetry $-\tau$ changes the sign of $H_3$, $F_0$ and $F_4$, on top of the pull-back, and preserves $F_2$: it is the orientifold symmetry $I_9\Omega$~\cite{Bergshoeff:2001pv}, with the worldsheet parity $\Omega$; this symmetry defines the O8-planes of type I$'$~\cite{Polchinski:1995df}. Then $\tau$ itself is $I_9\Omega(-1)^{F_L}$, under which $F_0$ is even (Sec.~\ref{sec:R7}).

Finally, the normalizations of the kinetic terms in \eqref{eq:kinred} relative to one another and to the Chern--Simons terms, which follow from the coefficients of \eqref{eq:S10} and are not fixed by the bulk (Sec.~\ref{sec:bdry}), are those which supersymmetry determines in type IIA.

\subsection{The equation of motion of the Romans mass}
\label{sec:zeromode}

In supergravity, $m$ is a parameter. In the formulation with a nine-form potential~\cite{Bergshoeff:2001pv}, it is already a field, constrained by ${\rm d} m=0$, but its constant value is not varied. Here, the variational principle of Sec.~\ref{sec:bdry} leaves all boundary values free, so that the constant mode of $A_0$, whose variation is, by \eqref{eq:puregauge} and \eqref{eq:HFsol}, $\delta A_0=\delta({\cal V}^{5/2})\,m+{\cal V}^{5/2}\,\delta m$, is varied together with the other fields. Its equation is contained in the Gauss law of $A_0$, ${\rm D} A_{10}=A_3A_8$. This is an eleven-form: it has no local content on $M_{10}$, and its content is global.

Suppose first that the holonomy is trivial, so that ${\cal V}$ and the forms $F_p$ of \eqref{eq:dict} are globally defined on $M_{11}$. The Gauss law then reads ${\rm d} F_{10}=H_3F_8$, and Stokes' theorem gives
\begin{equation}
\int_{M_{10}}F_{10}=\int_{M_{11}}H_3\,F_8\,.
\label{eq:zeromode}
\end{equation}
Conversely, since $H^{11}(M_{11},M_{10};\mathbb R)\cong\mathbb R$ for connected $M_{11}$, a ten-form on $M_{11}$ with a given exterior derivative and a given restriction to $M_{10}$ exists if and only if this condition holds. Equation~\eqref{eq:zeromode} is therefore the whole content of the Gauss law: it is the condition for $A_{10}$, whose boundary value is fixed by \eqref{eq:bcproj}, to extend over $M_{11}$.

When $B_2$ also extends over $M_{11}$, the right-hand side of \eqref{eq:zeromode} is a boundary term. By the Gauss laws ${\rm d} F_{2k}=H_3F_{2k-2}$ of the other even forms,
\begin{equation}
H_3F_8={\rm d}\Big[B_2F_8-\tfrac12\,B_2^2F_6+\tfrac16\,B_2^3F_4-\tfrac1{24}\,B_2^4F_2+\tfrac1{120}\,B_2^5F_0\Big]\,,
\label{eq:H3F8}
\end{equation}
and \eqref{eq:zeromode} becomes
\begin{equation}
\int_{M_{10}}\big(e^{-B_2}F\big)_{10}=0\,,\qquad F=\sum_{k=0}^{5}F_{2k}\,,
\label{eq:page}
\end{equation}
where the subscript denotes the component of degree ten. The polyform $e^{-B_2}F$ is closed by \eqref{eq:bianchi}, and its periods are Page charges~\cite{Marolf:2000cb}: the condition is that the Page charge of degree ten vanishes.

From the ten-dimensional point of view, \eqref{eq:page} is unexpected, since no equation arises from varying a parameter. Here, $m$ is varied, and \eqref{eq:page} is its equation of motion in the reduced action. To see this, let $\mathcal L$ be the full integrand of \eqref{eq:SR}, including the Chern--Simons terms, and differentiate it with respect to $m$ at fixed potentials. The result involves the Hodge duals of the electric field strengths, which the duality relations \eqref{eq:dual}, the equations of motion of the magnetic fields on $M_{10}$, convert into $F_6$, $F_8$ and $F_{10}$. Using also \eqref{eq:HFsol}, one finds
\begin{equation}
\frac{\partial\mathcal L}{\partial m}=i\,\Big[\big(e^{-B_2}F\big)_{10}-\tfrac{1}{24}\,{\rm d}\big(B_2^4\,C_1\big)\Big]\,.
\label{eq:dLdm}
\end{equation}
When the potentials are globally defined on $M_{10}$, the exact term integrates to zero, and \eqref{eq:page} becomes
\begin{equation}
\frac{\partial S_{\rm bTFT}}{\partial m}=0\,.
\label{eq:dSdm}
\end{equation}
In the reduction of Sec.~\ref{sec:reduction}, $m$ appeared as an integration constant; the eleven-form Gauss law is its equation of motion. The classical solutions of our action are thus the solutions of Romans' equations for which the action is also stationary with respect to $m$, a condition that has no counterpart in supergravity. In this respect, the Romans mass is a global dynamical variable, like the cosmological constant in the formulation of Ref.~\cite{Henneaux:1984ji}.

Three remarks are in order. First, for $m\neq0$, the condition \eqref{eq:dSdm} follows from the field equations. The action \eqref{eq:SR} is invariant under the rigid transformation \eqref{eq:rigid}, which rescales $m$ by $\lambda_0^{-5}$, and for globally defined potentials the corresponding Noether identity gives $m\,\partial S_{\rm bTFT}/\partial m=0$ on shell. Only at $m=0$ is \eqref{eq:page} a new condition: it constrains configurations of massless type IIA that supergravity would allow. The configurations excluded by \eqref{eq:page} are recovered if the pair $(A_0,A_{10})$ carries the Dirichlet condition $A_0=0$ on $M_{10}$ instead of the twisted condition (Sec.~\ref{sec:twist}). Since $A_0$ is then fixed, $m$ is not varied. The Gauss law still gives \eqref{eq:page}, but it now constrains only the boundary value of $A_{10}$, which is free. The result is massless type IIA as in supergravity. Second, when $B_2$ does not extend over $M_{11}$, the condition remains in the form \eqref{eq:zeromode}, which depends on the extension of $H_3$ and $F_8$, as the Chern--Simons term does (Sec.~\ref{sec:fromboundary}). Third, when the holonomy acts nontrivially on $A_0$, there is no covariantly constant section, the corresponding relative cohomology group vanishes, and the Gauss law imposes no condition; we show in Sec.~\ref{sec:D8} that there is then no constant mode to vary either.

These statements assume that $M_{10}$ is closed. Otherwise, the variational principle must be supplemented by conditions in the asymptotic region. If $A_0$ is held fixed there, its constant mode is not varied, and $m$ labels a sector of the theory, as in supergravity. Equation~\eqref{eq:zeromode} then acquires a contribution from the asymptotic region, and determines the flux of $A_{10}$ through it instead of constraining the fields on $M_{10}$.

\section{Holonomies and branes}
\label{sec:global}

Romans' action is recovered locally. Globally, our action also contains the $\mathrm{GL}(1,\mathbb{R})$ connection, which the Gauss laws require to be flat. Since the action is written in terms of this connection and the field strengths, rather than the dilaton and the potentials, its global structure is visible in the action itself: monodromies are holonomies of the flat connection, and branes enter as currents in the Gauss laws. In actions written in terms of potentials, such as the duality-symmetric actions of PST type~\cite{Bandos:2003et,DallAgata:1998va}, these features are not built in. We describe the holonomies (Sec.~\ref{sec:holonomy}), the constraints that the Romans mass, the fluxes and the branes impose on them (Sec.~\ref{sec:D8}), and the defects on which the hypersurface dual to the $\mathrm{O}(1)$ holonomy ends (Sec.~\ref{sec:R7}). The analysis is classical: the $\mathbb{R}^+$ holonomies rely on the rigid scaling \eqref{eq:rigid}, which the quantization of the fluxes is expected to break (Sec.~\ref{sec:discussion}).

\subsection{Wilson lines and holonomies}
\label{sec:holonomy}

Since $\mathrm{O}(1)$ is discrete, a $\mathrm{GL}(1,\mathbb{R})$ connection is a pair $(A_1,L)$ of a one-form and a real line bundle. On the overlap of two patches $U_i$ and $U_j$, the even forms are glued by signs $\sigma_{ij}=\pm1$, as in \eqref{eq:O1}, which are the transition functions of $L$. Being constant, they do not shift $A_1$, which is a globally defined one-form; $L$ enters only through the even forms, which are its sections. A gauge transformation \eqref{eq:O1} with locally constant signs $\sigma_i$ changes $\sigma_{ij}$ to $\sigma_i\,\sigma_{ij}\,\sigma_j$, and leaves invariant the first Stiefel--Whitney class $w_1(L)\in H^1(M_{11};\mathbb{Z}_2)$. Its evaluation $\langle w_1(L),\gamma\rangle\in\mathbb{Z}_2$ on a loop $\gamma$ is $1$ if the restriction of $L$ to $\gamma$ is a M\"obius band, and $0$ if it is a cylinder.

{Parallel transport along a curve $\gamma$ from a point $x$ to a point $y$ is the element}
\begin{equation}
\lambda_\gamma=\sigma_\gamma\,\exp\Big(\tfrac12\int_\gamma A_1\Big)\,,
\label{eq:holonomy}
\end{equation}
{where $\sigma_\gamma=\pm1$ is the product of the transition functions of $L$ met along $\gamma$; this sign cannot be written in terms of $A_1$, since $-1$ lies in the other component of $\mathrm{GL}(1,\mathbb{R})$. Under a gauge transformation \eqref{eq:Rgauge}, $\lambda_\gamma\to\lambda(y)\,\lambda_\gamma\,\lambda(x)^{-1}$. In the representation with weight $w$ and parity $\eta$, $\lambda_\gamma$ acts by the Wilson line}
\begin{equation}
W_{w,\eta}(\gamma)=|\lambda_\gamma|^{2w}\,\sigma_\gamma^{\eta}=\exp\Big(w\int_\gamma A_1\Big)\,\sigma_\gamma^{\eta}\,,
\label{eq:wilson}
\end{equation}
{which transports the zero-forms of this representation that are covariantly constant along $\gamma$.}

{Two fields have homogeneous Gauss laws, ${\rm D} A_0=0$ and ${\rm D} A_3=0$. The zero-form $A_0$ is covariantly constant along $\gamma$; for $A_3$, only its flux is determined along a curve. Let $\Sigma$ be a closed oriented three-dimensional submanifold on which the holonomy is trivial, so that $A_1$ can be gauged to zero on $\Sigma$. We carry $\Sigma$ along $\gamma$, with position $\Sigma_x$ at the point $x$, and set $\Phi_3(x)=\int_{\Sigma_x}A_3$ in this gauge. Since $A_1$ is flat, Stokes' theorem on the region swept gives ${\rm d}\Phi_3=w_3\,A_1\,\Phi_3$ along $\gamma$. With $\Phi_0=A_0$, both are transported by the Wilson line,}
\begin{equation}
\Phi_p(y)=W_p(\gamma)\,\Phi_p(x)\,,\qquad W_p(\gamma)\equiv W_{w_p,\eta_p}(\gamma)=\lambda_\gamma^{2w_p}\,,\qquad p=0,3\,,
\label{eq:holaction}
\end{equation}
{with $W_0(\gamma)=\lambda_\gamma^{5}$ and $W_3(\gamma)=|\lambda_\gamma|^{-2}$ in the compact notation of \eqref{eq:Rgauge}. The fluxes of the other forms are discussed in Sec.~\ref{sec:D8}.}

{For a closed curve, $\lambda_\gamma$ is the holonomy, and the Wilson loops are gauge invariant. We call $|\lambda_\gamma|$ and $\sigma_\gamma$ the $\mathbb{R}^+$ and $\mathrm{O}(1)$ holonomies. The first is fixed by the period of $A_1$ around $\gamma$, and the second by the topology of $L$, $\sigma_\gamma=(-1)^{\langle w_1(L),\gamma\rangle}$; it acts on each field as $\sigma_\gamma^{\eta_p}$, so that around a loop on which $L$ is a M\"obius band the even forms change sign and the odd forms do not. For a flat connection, the holonomy depends only on the homotopy class of $\gamma$, and the flat connections up to gauge are}
\begin{equation}
\mathrm{Hom}\big(\pi_1(M_{11}),\mathrm{GL}(1,\mathbb{R})\big)\cong H^1(M_{11};\mathbb R)\times H^1(M_{11};\mathbb{Z}_2)\,.
\label{eq:flatmod}
\end{equation}
{The periods of $A_1$ are a continuous modulus, and $w_1(L)$ a discrete one.}

In type IIA, a nontrivial $\mathbb{R}^+$ holonomy is a monodromy of the dilaton, which shifts by $\Delta\phi=4\log|\lambda_\gamma|$ around $\gamma$, as in the J-folds of type IIB, whose monodromy is hyperbolic~\cite{Inverso:2016eet,Assel:2018vtq,Joung:IIB}. A nontrivial $\mathrm{O}(1)$ holonomy is a $(-1)^{F_L}$ monodromy, under which the RR fields change sign: $\mathrm{O}(1)$ is generated by $(-1)^{F_L}$ (Sec.~\ref{sec:dictionary}), and $L$ is the $\mathbb{Z}_2$ gauge field of this symmetry. Such monodromies occur around a circle with a Wilson line for $(-1)^{F_L}$ in type IIA~\cite{Hellerman:2005ja}, and around the reflection seven-branes of Sec.~\ref{sec:R7}; in type IIB, the latter are holonomies of the reflections that extend $\mathrm{SL}(2,\mathbb{R})$ to $\mathrm{SL}^\pm(2,\mathbb R)$~\cite{Dierigl:2022reg,Joung:IIB}. The Poincar\'e dual of $w_1(L)$ is represented by a hypersurface $\mathcal Y^{\mathrm{O}(1)}$ across which the fields are glued by $(-1)^{F_L}$, and $\sigma_\gamma=(-1)^{\#(\gamma\cap\mathcal Y^{\mathrm{O}(1)})}$, where $\#(\gamma\cap\mathcal Y^{\mathrm{O}(1)})$ is the number of times $\gamma$ crosses $\mathcal Y^{\mathrm{O}(1)}$. This hypersurface is not a physical object: like a Dirac sheet, its position is a choice of gauge, and it cannot be detected locally, since every term of the action contains an even number of RR fields. Only its homology class is physical, or, if it ends in the interior of $M_{11}$, its boundary (Sec.~\ref{sec:R7}).

The holonomies \eqref{eq:flatmod} are those of loops in $M_{11}$, while the monodromies of type IIA are defined around loops in $M_{10}$. Since the connection is flat and abelian, a loop of $M_{10}$ that bounds a surface in $M_{11}$ has trivial holonomy, unless $\mathcal Y^{\mathrm{O}(1)}$ ends in the interior (Sec.~\ref{sec:R7}). For $M_{10}=S^1\times\mathbb R^9$, a monodromy around the circle is thus excluded if $M_{11}=D^2\times\mathbb R^9$, and allowed if $M_{11}=S^1\times\mathbb R^9\times\mathbb R_{\ge0}$. The fluxes are restricted in the same way: where the holonomy is trivial, $H_3$ is closed on $M_{11}$, and its flux vanishes through every cycle of $M_{10}$ that bounds in $M_{11}$. Classically, different fillings of the same boundary therefore allow different configurations on it; quantum mechanically, we expect to require that the result not depend on the filling, which should lead to quantization conditions (Sec.~\ref{sec:discussion}).

\subsection{The Romans mass, fluxes and branes}
\label{sec:D8}

Classically, the massless theory without fluxes allows any holonomy, and twisted circle reductions of massless type IIA by the dilaton scaling are known~\cite{Lavrinenko:1997qa,Bergshoeff:2002mb}. The Romans mass and the fluxes restrict it. Without sources, the Gauss law ${\rm D} A_0=0$ makes $A_0$ covariantly constant, so that, by \eqref{eq:holaction}, transport around a loop $\gamma$ multiplies it by $W_0(\gamma)$. Since $A_0$ is a section of the real line bundle $L$ of Sec.~\ref{sec:holonomy}, it must also return to its value. Hence
\begin{equation}
\big(W_0(\gamma)-1\big)\,A_0=0\,:
\label{eq:nohol}
\end{equation}
{a nontrivial holonomy forces $A_0=0$, that is, $m=0$, as a flat bundle admits a nonzero covariantly constant section only if its holonomy acts trivially on it.}

D8-branes relax this condition. Here we keep only what the holonomies require; a full discussion of the branes will be given in a forthcoming paper~\cite{Joung:inprep}. A D8-brane is a source for the Gauss law of $A_0$,
\begin{equation}
{\rm D} A_0=J_1\,,\qquad J_1=\sum_jq_j\,\delta_j\,.
\label{eq:D8source}
\end{equation}
Here $q_j$ is the charge of the $j$-th D8-brane, and $\delta_j$ is the Poincar\'e dual of a hypersurface $\mathcal Y_j\subset M_{11}$, a delta-function one-form supported on $\mathcal Y_j$ and normalized by $\int_{M_{11}}\omega\,\delta_j=\int_{\mathcal Y_j}\omega$ for any ten-form $\omega$. The hypersurface $\mathcal Y_j$ ends on the world-volume of the brane in $M_{10}$. The charges $q_j$ are sections of the line bundle $L$ over $\mathcal Y_j$ with the weight $\frac52$ of $A_0$, which makes \eqref{eq:D8source} covariant under \eqref{eq:Rgauge}. Consistency then requires them to be covariantly constant: since ${\rm D}^2=0$ and $\delta_j$ is closed in the interior of $M_{11}$, applying ${\rm D}$ to \eqref{eq:D8source} gives $\sum_j({\rm D}q_j)\,\delta_j=0$, that is,
\begin{equation}
{\rm D}q_j={\rm d}q_j-\tfrac52\,A_1\,q_j=0\quad\text{on }\mathcal Y_j\,,
\label{eq:qcovconst}
\end{equation}
with $A_1$ pulled back to $\mathcal Y_j$. On the local solution of Sec.~\ref{sec:solve}, $q_j={\cal V}^{5/2}n_j$, where $n_j$ is the D8-brane charge~\cite{Polchinski:1995mt,Bergshoeff:1995super8}, and \eqref{eq:qcovconst} states that $n_j$ is constant along the brane.

The source arises from the coupling of $A_{10}$ to $\mathcal Y_j$,
\begin{equation}
S_{\rm D8}=i\sum_j\int_{\mathcal Y_j}q_j\,A_{10}=i\int_{M_{11}}A_{10}\,J_1\,,
\label{eq:D8coupling}
\end{equation}
whose variation with respect to $A_{10}$, added to that of the term $-i\,A_{10}\,{\rm D} A_0$ of \eqref{eq:SBF}, gives \eqref{eq:D8source}; it is neutral, since $q_j$ carries the weight $\frac52$.

The other Gauss laws constrain the source further. Applying ${\rm D}$ to ${\rm D} A_2=A_3A_0$ gives $A_3J_1=0$: either $A_3$ pulls back to zero on $\mathcal Y_j$, or the higher even forms are sourced as well, ${\rm D}_HA=J$ with ${\rm D}_H={\rm D}-A_3$ and $A=\sum_kA_{2k}$, by a current with ${\rm D}_HJ=0$. Such a current is $J=\sum_jq_j\,e^{\mathcal F_j}\delta_j$, where the world-volume flux $\mathcal F_j$ of the D8-brane is a two-form on $\mathcal Y_j$ with the weight of $A_3$ and ${\rm D}\mathcal F_j=A_3$. It also makes the coupling independent of the auxiliary hypersurfaces. By ${\rm D} A_{10}=A_3A_8$, a deformation of $\mathcal Y_j$ at fixed boundary changes \eqref{eq:D8coupling} by the integral of $q_jA_3A_8$ over the region swept, and the terms in $\mathcal F_j$ compensate this change: $J$ corresponds to the coupling $i\sum_j\int_{\mathcal Y_j}q_j\,\big(e^{-\mathcal F_j}A\big)_{10}$, with the signs of the kinetic terms in \eqref{eq:S11}, whose integrand is closed wherever $\mathcal F_j$ extends. Finally, applying ${\rm d}$ to the Gauss law of $A_9$ leaves terms such as $\tfrac52J_1A_{10}$, so that this Gauss law must in general be sourced as well, by a current coupled to $A_1$. These sources, the world-volume fields and the global form of these statements are left to that paper; the argument below uses only \eqref{eq:D8source}.

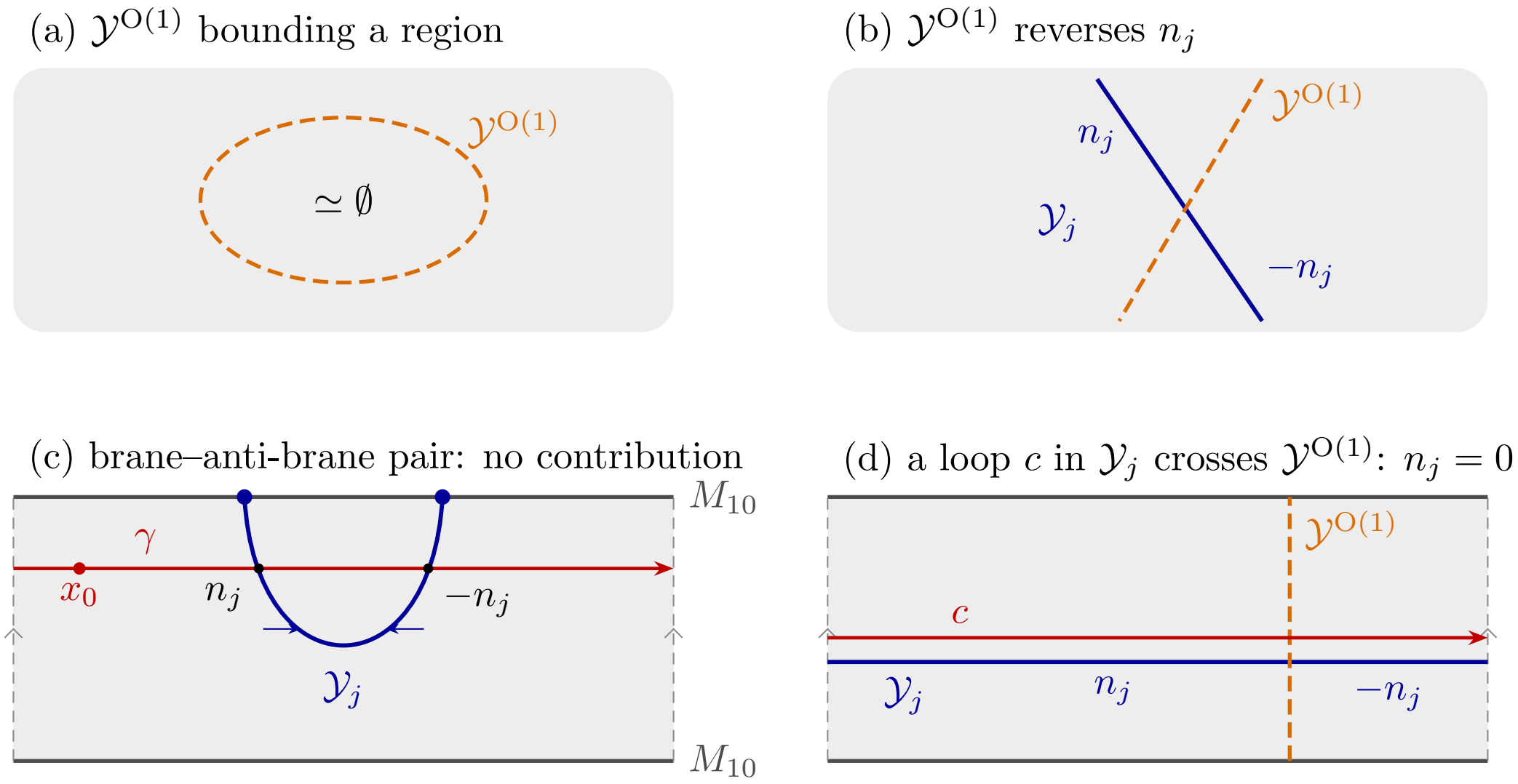
\begin{figure}[t]
\centering
\begin{tikzpicture}[
  lab/.style={font=\small},
  sub/.style={font=\small, anchor=west},
  Yl/.style={blue!60!black, line width=1.1pt},
  Wl2/.style={orange!85!black, line width=1pt, dash pattern=on 4pt off 2pt},
  gam2/.style={red!75!black, line width=0.9pt, -{Stealth[length=5pt]}},
  gam2b/.style={red!55, line width=0.9pt, -{Stealth[length=5pt]}},
  D8pt/.style={blue!60!black},
  patch/.pic={
    \fill[black!7, rounded corners=8pt] (0,0) rectangle (6,2.4);
  },
  frame/.pic={
    \fill[black!7] (0,0) rectangle (6,2.4);
    \draw[black!70, line width=1pt] (0,0) -- (6,0) (0,2.4) -- (6,2.4);
    \draw[black!45, densely dashed] (0,0) -- (0,2.4) (6,0) -- (6,2.4);
    \foreach \x in {0,6} \draw[black!45] (\x-0.08,1.12) -- (\x,1.2) -- (\x+0.08,1.12);
  }]
\begin{scope}[shift={(0,3.6)}]
\pic {patch};
\node[sub] at (0,2.75) {(a) $\mathcal Y^{\mathrm{O}(1)}$ bounding a region};
\draw[Wl2] (3.0,1.2) ellipse (1.3 and 0.75);
\node[lab, orange!85!black] at (4.55,1.85) {$\mathcal Y^{\mathrm{O}(1)}$};
\node[lab] at (3.0,1.2) {$\simeq\emptyset$};
\end{scope}
\begin{scope}[shift={(7.4,3.6)}]
\pic {patch};
\node[sub] at (0,2.75) {(b) $\mathcal Y^{\mathrm{O}(1)}$ reverses $n_j$};
\draw[Yl] (2.45,2.3) -- (3.95,0.1);
\draw[Wl2] (3.95,2.3) -- (2.65,0.1);
\node[lab, orange!85!black, right] at (3.9,2.1) {$\mathcal Y^{\mathrm{O}(1)}$};
\node[lab, blue!60!black] at (2.45,1.75) {$n_j$};
\node[lab, blue!60!black] at (4.3,0.55) {$-n_j$};
\node[lab, blue!60!black, below] at (2.1,1.3) {$\mathcal Y_j$};
\end{scope}
\begin{scope}[shift={(0,-0.3)}]
\pic {frame};
\node[sub] at (0,2.75) {(c) brane--anti-brane pair: no contribution};
\draw[Yl] (2.1,2.4) .. controls (2.2,0.6) and (3.8,0.6) .. (3.9,2.4);
\fill[D8pt] (2.1,2.4) circle (2pt) (3.9,2.4) circle (2pt);
\node[lab, blue!60!black, below] at (3.0,0.95) {$\mathcal Y_j$};
\draw[blue!60!black, -{Stealth[length=3.5pt]}] (2.27,1.2) -- (2.62,1.2);
\draw[blue!60!black, -{Stealth[length=3.5pt]}] (3.73,1.2) -- (3.38,1.2);
\draw[gam2] (0,1.75) -- (6,1.75);
\fill[red!75!black] (0.6,1.75) circle (1.6pt) node[below, lab] {$x_0$};
\node[lab, red!75!black, above] at (1.2,1.75) {$\gamma$};
\fill (2.23,1.75) circle (1.3pt) node[below left=0pt, lab] {$n_j$};
\fill (3.77,1.75) circle (1.3pt) node[below right=0pt, lab] {$-n_j$};
\node[lab, black!70, right] at (6,2.4) {$M_{10}$};
\node[lab, black!70, right] at (6,0) {$M_{10}$};
\end{scope}
\begin{scope}[shift={(7.4,-0.3)}]
\pic {frame};
\node[sub] at (0,2.75) {(d) a loop $c$ in $\mathcal Y_j$ crosses $\mathcal Y^{\mathrm{O}(1)}$: $n_j=0$};
\draw[Yl] (0,0.9) -- (6,0.9);
\node[lab, blue!60!black, below] at (0.7,0.9) {$\mathcal Y_j$};
\draw[Wl2] (4.2,0) -- (4.2,2.4);
\node[lab, orange!85!black, right] at (4.2,2.1) {$\mathcal Y^{\mathrm{O}(1)}$};
\node[lab, blue!60!black, below] at (2.6,0.9) {$n_j$};
\node[lab, blue!60!black, below] at (5.1,0.9) {$-n_j$};
\draw[gam2] (0,1.12) -- (6,1.12);
\node[lab, red!75!black, above] at (1.2,1.12) {$c$};
\end{scope}
\end{tikzpicture}
\caption{D8-branes and the hypersurface $\mathcal Y^{\mathrm{O}(1)}$. Blue lines are hypersurfaces $\mathcal Y_j$, blue dots D8-branes, orange dashed lines $\mathcal Y^{\mathrm{O}(1)}$, and red lines loops. The upper panels are local in the bulk and show properties of $\mathcal Y^{\mathrm{O}(1)}$: (a) a component of $\mathcal Y^{\mathrm{O}(1)}$ that bounds a region can be removed; (b) the charge of $\mathcal Y_j$ changes sign across $\mathcal Y^{\mathrm{O}(1)}$, so that moving $\mathcal Y^{\mathrm{O}(1)}$ across $\mathcal Y_j$ reverses $n_j$. The lower panels show, in a cross-section of a solid torus like that of Figure~\ref{fig:D8loop}, configurations that are set aside; the horizontal direction is the non-contractible cycle, with the left and right edges identified, and the top and bottom edges lie on $M_{10}$. (c) A hypersurface that ends on a brane--anti-brane pair is crossed in both directions relative to its orientation (arrows), and does not contribute to $N$. (d) If $\mathcal Y_j$ contains a loop $c$ that crosses $\mathcal Y^{\mathrm{O}(1)}$ once, $n_j$ comes back as $-n_j$, and the charge vanishes.}
\label{fig:D8cases}
\end{figure}

\begin{figure}[t]
\centering
\includegraphics{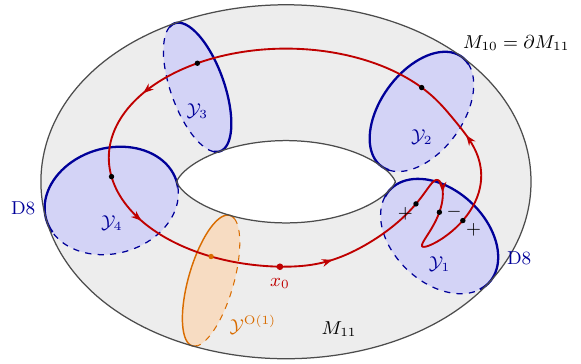}
\caption{D8-branes and the curve $\gamma$, schematically. The solid torus is $M_{11}$, and its surface is $M_{10}$. The discs $\mathcal Y_j$ are the hypersurfaces of \eqref{eq:D8source}, and their boundaries on $M_{10}$ are the D8-branes. The curve $\gamma$ is a loop based at $x_0$, around the non-contractible cycle of $M_{11}$. It crosses $\mathcal Y_1$ three times, the second time against the orientation, so that these crossings contribute $n_1-n_1+n_1=n_1$ to $N$ in \eqref{eq:D8charge}, and each of the other discs once. The orange disc is the hypersurface $\mathcal Y^{\mathrm{O}(1)}$ of Sec.~\ref{sec:holonomy}, across which the fields are glued by $(-1)^{F_L}$. It is placed just before $x_0$, so that $\gamma$ crosses it once, at the end of the turn, and $\sigma_\gamma=-1$. Following $F_0$ once around $\gamma$ gives \eqref{eq:D8charge}.}
\label{fig:D8loop}
\end{figure}

To see how D8-branes relax \eqref{eq:nohol}, we follow the Romans mass around a loop. Along a curve from a base point $x_0$, ${\cal V}(x)={\cal V}(x_0)\exp\int_{x_0}^xA_1$, and the local solution of Sec.~\ref{sec:solve} gives $A_0={\cal V}^{5/2}F_0$ and $q_j={\cal V}^{5/2}n_j$. The dilaton factors then drop out, and $F_0$ obeys three rules: (i) it is constant away from the hypersurfaces; (ii) it jumps by $n_j$ across $\mathcal Y_j$; (iii) it changes sign across the hypersurface $\mathcal Y^{\mathrm{O}(1)}$ of Sec.~\ref{sec:holonomy}, where the frame of the line bundle $L$ is reversed.

As noted in Sec.~\ref{sec:holonomy}, only the parity of the number of crossings of $\mathcal Y^{\mathrm{O}(1)}$ is physical: $\mathcal Y^{\mathrm{O}(1)}$ can be moved by $\mathrm{O}(1)$ gauge transformations, and a component that bounds a region, such as the region between two parallel components, is removed by the transformation $\sigma=-1$ on that region (Figure~\ref{fig:D8cases}a). By rule (iii), the charge of $\mathcal Y_j$ changes sign across $\mathcal Y^{\mathrm{O}(1)}$ (Figure~\ref{fig:D8cases}b): moving $\mathcal Y^{\mathrm{O}(1)}$ across $\mathcal Y_j$ reverses $n_j$, so that a brane on one side of $\mathcal Y^{\mathrm{O}(1)}$ is an anti-brane on the other, and where $\mathcal Y_j$ meets $\mathcal Y^{\mathrm{O}(1)}$, its charge changes sign along $\mathcal Y_j$. When $\mathcal Y^{\mathrm{O}(1)}$ can be moved off $\mathcal Y_j$, this is a choice of gauge.

Two kinds of configurations can be set aside. A hypersurface that ends on a brane--anti-brane pair, or on a D8-brane on a contractible or separating cycle, is crossed by every loop equally often in both directions, and never contributes (Figure~\ref{fig:D8cases}c). The intersection of $\mathcal Y_j$ with $\mathcal Y^{\mathrm{O}(1)}$ cannot be removed if the connected hypersurface $\mathcal Y_j$ contains a loop $c$ that crosses $\mathcal Y^{\mathrm{O}(1)}$ an odd number of times (Figure~\ref{fig:D8cases}d). Following $\mathcal Y_j$ around $c$, $n_j$ then comes back as $-n_j$, so that $n_j=0$. The same holds if the dilaton has a monodromy along $c$: $n_j$ is constant along the brane, but $q_j={\cal V}^{5/2}n_j$ must return to its value while ${\cal V}$ does not. Both cases follow from $\big(W_0(c)-1\big)\,q_j=0$, as \eqref{eq:nohol} did. Since the world-volume of the brane lies in $\mathcal Y_j$, a D8-brane cannot wrap a cycle around which the dilaton or $(-1)^{F_L}$ has a monodromy; only the latter is expected to survive in the quantum theory (Sec.~\ref{sec:discussion}).

We now follow $F_0$ around a loop $\gamma$ based at $x_0$ and transverse to the hypersurfaces and to $\mathcal Y^{\mathrm{O}(1)}$ (Figure~\ref{fig:D8loop}). At each crossing of $\mathcal Y_j$, $F_0$ changes by $n_j$ or $-n_j$, according to the orientation of the crossing. By the rules above, we place $\mathcal Y^{\mathrm{O}(1)}$ away from the charged hypersurfaces, so that $\gamma$ crosses it at most once, just before it returns to $x_0$. After one turn,
\begin{equation}
F_0'(x_0)=\sigma_\gamma\big(m_0+N\big)\,,\qquad {N=\sum_k\epsilon_k\,n_{j_k}}\,,\qquad m_0=F_0(x_0)\,,
\label{eq:D8charge}
\end{equation}
where $k$ labels the crossings of $\gamma$ with the hypersurfaces, $\mathcal Y_{j_k}$ is the hypersurface crossed at the $k$-th crossing, and $\epsilon_k=\pm1$ is the sign of that crossing relative to the orientation of $\mathcal Y_{j_k}$. Further, $\sigma_\gamma=(-1)^{\langle w_1(L),\gamma\rangle}$ is $-1$ if $\gamma$ crosses $\mathcal Y^{\mathrm{O}(1)}$ and $1$ otherwise. With $\mathcal Y^{\mathrm{O}(1)}$ placed elsewhere, the branes beyond it would count as anti-branes, with the same $N$.

Closing the loop brings in the dilaton: after one turn, ${\cal V}(x_0)$ has become ${\cal V}'(x_0)={\cal V}(x_0)\exp\oint_\gamma A_1$. Since $A_0={\cal V}^{5/2}F_0$ is single-valued, ${\cal V}'(x_0)^{5/2}F_0'(x_0)={\cal V}(x_0)^{5/2}m_0$, that is,
\begin{equation}
m_0=W_0(\gamma)\,\big(m_0+N\big)\,,\qquad
W_0(\gamma)=\sigma_\gamma\Big(\frac{{\cal V}'(x_0)}{{\cal V}(x_0)}\Big)^{\frac52}\,,
\label{eq:transport}
\end{equation}
where $W_0(\gamma)$ is the holonomy \eqref{eq:holaction}. For $N=0$, this is \eqref{eq:nohol}. For $m_0+N\neq0$, it gives
\begin{equation}
W_0(\gamma)=\frac{m_0}{m_0+N}\,.
\label{eq:D8hol}
\end{equation}

For $m_0\neq0$, the holonomy is thus fixed by the D8-brane charges, including its $\mathrm{O}(1)$ part: $\sigma_\gamma=-1$ precisely when the Romans mass comes back with the opposite sign, and all RR fields are then reversed together around $\gamma$. This is possible only because $\mathrm{O}(1)$ is gauged; with the gauge group $\mathbb{R}^+$, only $\sigma_\gamma=1$ would be allowed. The base point enters only through $m_0$ and $N$, which change together when $x_0$ is moved, and for several independent loops it suffices to impose \eqref{eq:transport} on a set of generators of $H_1(M_{11})$.

Equation~\eqref{eq:D8hol} presupposes $m_0\neq0$ and $m_0+N\neq0$. If $m_0=0$, \eqref{eq:transport} gives $N=0$ for any holonomy, so that a Romans mass that returns to zero from $m_0\neq0$ is excluded. Conversely, $N\neq0$ requires $W_0(\gamma)\neq1$, which fixes the Romans mass, $m_0=N\,W_0(\gamma)/\big(1-W_0(\gamma)\big)$; the solution of \eqref{eq:D8source} is then unique, with no constant mode left to vary, as anticipated in Sec.~\ref{sec:zeromode}. For a loop that bounds a surface in $M_{11}$, the holonomy is trivial (Sec.~\ref{sec:holonomy}; see Sec.~\ref{sec:R7} for the exception), and $N=0$: the familiar vanishing of the total charge on a closed manifold is the case of trivial holonomy. For instance, let $M_{10}=S^1\times M_9$, with D8-branes wrapping $M_9$ at points $\theta_j$ of the circle. For $M_{11}=D^2\times M_9$, the circle bounds a disc, and $N=0$. For $M_{11}=S^1\times X_{10}$, with $\partial X_{10}=M_9$, each brane bounds $\{\theta_j\}\times X_{10}$, and $N\neq0$ is allowed, with the holonomy around the circle fixed by \eqref{eq:D8hol}; Figure~\ref{fig:D8loop} is the case $M_9=S^1$, $X_{10}=D^2$. Branes on separating cycles, on the other hand, never contribute to $N$, so that the sum of all the charges is not constrained (Figure~\ref{fig:D8cases}c).

In type IIA terms, \eqref{eq:D8hol} states that $A_0=e^{5\phi/4}F_0$ returns to its value around $\gamma$, although neither factor does: the jumps of the Romans mass across the D8-branes compensate the monodromy of the dilaton, $e^{5\Delta\phi/4}=m_0/(m_0+N)$ for $\sigma_\gamma=1$. Equivalently, the Romans potential, proportional to $e^{5\phi/2}m^2=A_0^2$, is single-valued, a form of the relation that does not depend on the normalization of $A_1$ and $\lambda$.

The relation \eqref{eq:D8hol} is classical. A nontrivial $\mathbb{R}^+$ holonomy rescales the fluxes through \eqref{eq:rigid}, which is incompatible with their quantization (Sec.~\ref{sec:discussion}). In the quantum theory, only $W_0(\gamma)=\pm1$ can remain, that is, $N=0$ or $N=-2m_0$: a nonzero total charge requires a $(-1)^{F_L}$ monodromy, and fixes $m_0=-N/2$. For two D8-branes of unit charge on a circle, for instance, the Romans mass is $-1$, $0$ and $1$ in the three regions into which the branes and $\mathcal Y^{\mathrm{O}(1)}$ divide the circle, and $\mathcal Y^{\mathrm{O}(1)}$ identifies $1$ with $-1$. Without D8-branes, a $(-1)^{F_L}$ monodromy requires $m_0=0$, as for the circle with a Wilson line for $(-1)^{F_L}$~\cite{Hellerman:2005ja}.

In the massless theory, \eqref{eq:nohol} is empty, but the three-form, the other field with a homogeneous Gauss law, obeys a similar constraint. (In the massive theory, this constraint is automatic: without sources, the Gauss law of $A_2$ gives $A_0A_3={\rm D}A_2$, the counterpart of $mH_3={\rm d}F_2$ in Romans' theory, so that $\Phi_3$ vanishes wherever $A_0\neq0$.) The flux $\Phi_3$ of $A_3$ through a closed three-cycle $\Sigma$ is transported by \eqref{eq:holaction} with $W_3(\gamma)=|\lambda_\gamma|^{-2}$. This is not a change of the flux of $H_3$: on the local solution \eqref{eq:dict}, $\Phi_3(x)={\cal V}(x)^{-1}\,h$, where $h=\int_{\Sigma_x}H_3$ does not depend on $x$, and the Wilson line is the ratio of the dilaton factors at the two ends of $\gamma$. A condition arises only for a closed curve with a nontrivial holonomy. If $\Sigma$ returns to itself after one turn, with its orientation, the flux of $A_3$, which is unchanged by the higher-form gauge transformations that may glue $A_3$ across patches, must return to its value, while \eqref{eq:holaction} multiplies it by $W_3(\gamma)$. Hence
\begin{equation}
\big(W_3(\gamma)-1\big)\,\Phi_3=0\,.
\label{eq:noflux}
\end{equation}
For example, on $S^1\times\Sigma$ with $A_1=\alpha\,{\rm d}\theta$, so that the dilaton $\phi=2\alpha\theta$ is not periodic, a constant flux $H_3=h\,{\rm vol}_\Sigma$, with $\int_\Sigma{\rm vol}_\Sigma=1$, corresponds to $A_3=h\,e^{-\alpha\theta}\,{\rm vol}_\Sigma$. This form satisfies ${\rm D} A_3=0$ but is not periodic, and no higher-form gauge transformation can glue its values at $\theta=0$ and $\theta=2\pi$, since their difference has a nonzero integral over $\Sigma$. Hence $h=0$, just as \eqref{eq:nohol} requires $m=0$. In type IIA terms, a monodromy of the dilaton excludes $H_3$ fluxes through cycles that return to themselves, whereas a $(-1)^{F_L}$ monodromy, under which $H_3$ is even, does not.

NS5-branes can be included in the same way for the three-form, although less completely. Their world-volume theory, whose anomaly involves $X_8$ (Sec.~\ref{sec:democratic}), is beyond the scope of this paper, and for $m\neq0$ they cannot appear alone: an NS5-brane is a source ${\rm D} A_3=J_4=\sum_jq_j\,\delta_j$, and applying ${\rm D}$ to ${\rm D} A_2=A_3A_0$ gives $J_4A_0=0$, so that the Gauss law of $A_2$ must be sourced as well, by D6-branes ending on the NS5-branes~\cite{Hanany:1997sa}. With this proviso, the source arises from the coupling $-i\sum_j\int_{\mathcal Y_j}q_j\,A_7$, where $\mathcal Y_j\subset M_{11}$ is now seven-dimensional and ends on the world-volume of the brane. The charges have the weight $-1$ of $A_3$ and no parity, so that NS5-branes, unlike D8-branes, are not reversed by the $\mathrm{O}(1)$ holonomy, and their hypersurfaces may cross $\mathcal Y^{\mathrm{O}(1)}$ freely. The flux $\Phi_3$ jumps by $q_j$ wherever the region swept by $\Sigma$ meets $\mathcal Y_j$, and the argument that led to \eqref{eq:D8hol} gives
\begin{equation}
W_3(\gamma)=\frac{h_0}{h_0+N_5}\,,
\label{eq:NS5hol}
\end{equation}
where $h_0$ is the flux of $H_3$ through $\Sigma$ at the base point and $N_5$ is the total NS5-brane charge met. Since $W_3(\gamma)>0$, $h_0$ and $h_0+N_5$ have the same sign. In type IIA terms, the change of the flux of $H_3$ across the NS5-branes compensates the monodromy of the dilaton, $e^{-\Delta\phi/2}=h_0/(h_0+N_5)$. Without monodromy, \eqref{eq:NS5hol} is the vanishing of the total NS5-brane charge on the closed four-manifold swept by $\Sigma$, the only case that can remain in the quantum theory.

For the RR forms, whose Gauss laws ${\rm D} A_{2k}=A_3A_{2k-2}$ are inhomogeneous, the flux through a closed $2k$-dimensional $\Sigma$ also changes by the integral of $A_3A_{2k-2}$ over the region swept. Where this term vanishes, as for the first nonvanishing form of the chain, the flux is transported by $W_{2k}(\gamma)$, and \eqref{eq:noflux} holds with $\Phi_{2k}$ in place of $\Phi_3$. Since $\eta_{2k}=1$, such a flux is then excluded by a $(-1)^{F_L}$ monodromy as well as by a monodromy of the dilaton. In general, the same holds for the Page combination $\big(e^{-\beta}A\big)_{2k}$, with ${\rm D}\beta=A_3$ (Sec.~\ref{sec:zeromode}). These statements assume that there are no brane sources, which would source the Gauss laws of the RR forms as well; the relations \eqref{eq:D8hol} and \eqref{eq:NS5hol} involve only the sources of $A_0$ and $A_3$, and are not affected.

\subsection{The ends of $\mathcal Y^{\mathrm{O}(1)}$: reflection seven-branes}
\label{sec:R7}

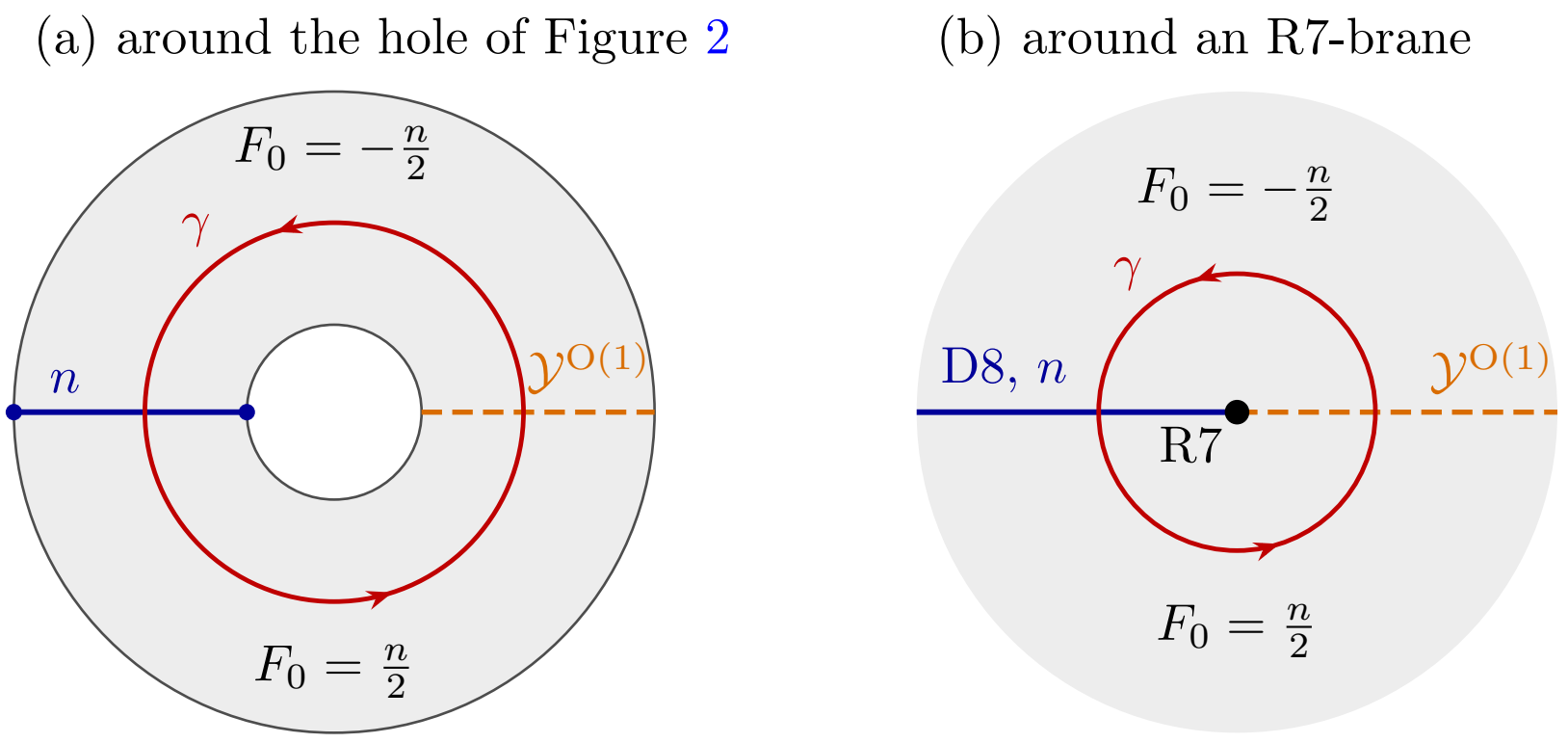
\begin{figure}[t]
\centering
\begin{tikzpicture}[
  lab/.style={font=\small},
  sub/.style={font=\small, anchor=west},
  Yl/.style={blue!60!black, line width=1.1pt},
  Wl3/.style={orange!85!black, line width=1pt, dash pattern=on 4pt off 2pt},
  gam3/.style={red!75!black, line width=0.9pt,
    decoration={markings, mark=at position 0.3 with {\arrow{Stealth[length=5pt]}},
                          mark=at position 0.8 with {\arrow{Stealth[length=5pt]}}},
    postaction={decorate}},
  R7/.style={black}]
\begin{scope}
\fill[black!7, even odd rule] (0,0) circle (2.2) (0,0) circle (0.6);
\draw[black!70, line width=0.5pt] (0,0) circle (2.2) (0,0) circle (0.6);
\node[sub] at (-2.2,2.55) {(a) around the hole of Figure~\ref{fig:D8loop}};
\draw[Wl3] (0.6,0) -- (2.2,0);
\node[lab, orange!85!black, above] at (1.75,0) {$\mathcal Y^{\mathrm{O}(1)}$};
\draw[Yl] (-0.6,0) -- (-2.2,0);
\fill[blue!60!black] (-0.6,0) circle (1.6pt) (-2.2,0) circle (1.6pt);
\node[lab, blue!60!black, above] at (-1.85,0) {$n$};
\draw[gam3] (0,0) circle (1.3);
\node[lab, red!75!black] at (-0.95,1.25) {$\gamma$};
\node[lab] at (0,1.78) {$F_0=-\tfrac n2$};
\node[lab] at (0,-1.78) {$F_0=\tfrac n2$};
\end{scope}
\begin{scope}[shift={(6.2,0)}]
\fill[black!7] (0,0) circle (2.2);
\node[sub] at (-2.2,2.55) {(b) around an R7-brane};
\draw[Wl3] (0,0) -- (2.2,0);
\node[lab, orange!85!black, above] at (1.75,0) {$\mathcal Y^{\mathrm{O}(1)}$};
\draw[Yl] (0,0) -- (-2.2,0);
\fill[blue!60!black] (-2.2,0) circle (0pt);
\node[lab, blue!60!black, above] at (-1.6,0) {D8, $n$};
\draw[gam3] (0,0) circle (0.95);
\fill[R7] (0,0) circle (2.4pt);
\node[lab, below left=-1pt] at (0,0) {R7};
\node[lab, red!75!black] at (-0.75,0.95) {$\gamma$};
\node[lab] at (0,1.5) {$F_0=-\tfrac n2$};
\node[lab] at (0,-1.5) {$F_0=\tfrac n2$};
\end{scope}
\end{tikzpicture}
\caption{The R7-brane in place of a non-contractible cycle. (a) A section of the solid torus of Figure~\ref{fig:D8loop} along its core, with a single hypersurface $\mathcal Y_j$ of charge $n$ (blue), whose ends on $M_{10}$ are D8-branes, and the hypersurface $\mathcal Y^{\mathrm{O}(1)}$ (orange); the loop $\gamma$ goes around the hole. (b) The plane transverse to an R7-brane in $M_{10}$, with a D8-brane of charge $n$ ending on it; the loop $\gamma$ links the brane, which is the end of $\mathcal Y^{\mathrm{O}(1)}$. In both cases, $\gamma$ crosses the D8-brane hypersurface and $\mathcal Y^{\mathrm{O}(1)}$ once, and with trivial $\mathbb R^+$ holonomy the Romans mass is $-n/2$ and $n/2$ on the two sides, values identified across $\mathcal Y^{\mathrm{O}(1)}$. For $n=0$, it vanishes. Panel (b) is the limit of panel (a) in which the hole shrinks to a point.}
\label{fig:R7}
\end{figure}

So far, $L$ has been defined on all of $M_{11}$, so that $\mathcal Y^{\mathrm{O}(1)}$ does not end in the interior. This can be relaxed by letting $\mathcal Y^{\mathrm{O}(1)}$ end on a submanifold $\mathcal R\subset M_{11}$ of codimension two, around which the $\mathrm{O}(1)$ holonomy is $\sigma=-1$. A small loop around $\mathcal R$ then behaves as the loop around the non-contractible cycle of Figure~\ref{fig:D8loop}, and $\mathcal R$ can be viewed as the limit of that cycle in which its hole shrinks to a locus of codimension two (Figure~\ref{fig:R7}). The results of Sec.~\ref{sec:D8} therefore carry over, and we describe only what is new. These loci, like the hypersurfaces of the D8-branes, will be treated in general in the forthcoming paper~\cite{Joung:inprep} mentioned in Sec.~\ref{sec:D8}. The hypersurface $\mathcal Y^{\mathrm{O}(1)}$ and the locus $\mathcal R$ are of different nature. The position of $\mathcal Y^{\mathrm{O}(1)}$ is a choice of frame for $L$, which an $\mathrm{O}(1)$ gauge transformation changes, as for a Dirac sheet. A loop around $\mathcal R$, by contrast, crosses $\mathcal Y^{\mathrm{O}(1)}$ once wherever it is placed, so that $\mathcal R$ cannot be removed, as a monopole cannot. Its position in the interior is nevertheless unobservable, since the bulk is topological, and only its end on $M_{10}$, the world-volume of the R7-brane discussed below, is fixed, as for the hypersurfaces $\mathcal Y_j$ and the D8-branes on which they end. The locus $\mathcal R$ is not a source in the Gauss laws, and $A_1$ remains flat. It is instead a magnetic source for $L$, as a monopole is for a $\mathrm U(1)$ gauge field. The $\mathrm{O}(1)$ holonomy of a loop is measured by the delta-form $\delta_{\mathcal Y^{\mathrm{O}(1)}}$ of the hypersurface, which counts its crossings and represents $w_1(L)$. Since $\mathcal Y^{\mathrm{O}(1)}$ ends on $\mathcal R$, this form is no longer closed: ${\rm d}\delta_{\mathcal Y^{\mathrm{O}(1)}}=\delta_{\mathcal R}$ mod $2$, just as the Bianchi identity ${\rm d}F=0$ fails at a monopole.

Where $\mathcal R$ meets $M_{10}$, it leaves the world-volume of a seven-brane around which the RR fields return with the opposite sign, the reflection seven-brane, or R7-brane, of $(-1)^{F_L}$. R7-branes were first proposed in type IIB, as non-supersymmetric objects required by the cobordism conjecture~\cite{McNamara:2019rup,Dierigl:2022reg,Debray:2023yrs}, and the $(-1)^{F_L}$ R7-brane exists in type IIA as well, as a vortex of the $\mathbb Z_2$ gauge field of $(-1)^{F_L}$~\cite{Heckman:2025gso,Chakrabhavi:2025pintopia}.\footnote{In type IIB, R7-branes come with monodromy $\Omega$ or $(-1)^{F_L}$; strings and D3-branes can end on them~\cite{Dierigl:2022reg}, and, wrapped at infinity, they act as charge-conjugation operators~\cite{Dierigl:2023jdp}. In type IIA, the R7-brane arises from a collapsing IIA/IIB wall~\cite{Heckman:2025gso}, D$p$-branes can end on it, and in M-theory it reverses the eleventh circle~\cite{Chakrabhavi:2025pintopia}. Its observability is discussed in Ref.~\cite{Sen:2025exotic}.} Here this gauge field is the line bundle $L$, and the topological bulk also contains the Romans mass and the D8-branes, which leads to the constraints derived below. The core of the R7-brane, its tension and its world-volume degrees of freedom are beyond the bulk theory, as for the D8-brane, but the configurations around it are not.

The data of Sec.~\ref{sec:holonomy} change accordingly. The Wilson lines are now defined for curves that avoid $\mathcal R$, and the $\mathrm{O}(1)$ holonomy is still $\sigma_\gamma=(-1)^{\#(\gamma\cap\mathcal Y^{\mathrm{O}(1)})}$, but a loop that bounds in $M_{11}$ no longer has trivial holonomy if it links $\mathcal R$. The locus $\mathcal R$ need not reach $M_{10}$, and it has an effect even when no R7-brane appears on the boundary. Take $M_{10}=S^1\times M_9$. If it is filled by $M_{11}=D^2\times M_9$, the circle bounds a disc, and Sec.~\ref{sec:holonomy} excluded a $(-1)^{F_L}$ monodromy around it. A locus $\mathcal R=\{p\}\times M_9$ at a point $p$ of the disc, entirely in the interior, lifts this restriction: the circle now links $\mathcal R$, and the monodromy is allowed. In effect, removing $p$ turns the disc into an annulus, and the filling $D^2\times M_9$ behaves as $S^1\times M_9\times\mathbb R_{\ge0}$. The fields are constrained as well. The even forms are sections of $L$, which changes sign around $\mathcal R$, so that, where they are continuous, they vanish there:
\begin{equation}
A_{2k}\big|_\mathcal R=0\,.
\label{eq:R7vanish}
\end{equation}

Around a small loop $\gamma$ linking the brane, the $\mathbb{R}^+$ holonomy is trivial when $A_1$ is regular near $\mathcal R$, so that $W_0(\gamma)=-1$, and \eqref{eq:transport} gives
\begin{equation}
N=-2m_0\,,
\label{eq:R7}
\end{equation}
the relation found in Sec.~\ref{sec:D8} for a $(-1)^{F_L}$ monodromy, now on loops that bound in $M_{11}$. Without D8-branes, the Romans mass therefore vanishes near the R7-brane, in agreement with \eqref{eq:R7vanish}. A D8-brane can, however, end on the R7-brane, if the Romans mass is $-n_j/2$ and $n_j/2$ on its two sides, values identified across $\mathcal Y^{\mathrm{O}(1)}$ (Figure~\ref{fig:R7}b): the D8-brane that ended on the inner boundary of Figure~\ref{fig:R7}a now ends on the R7-brane. Locally, the jump of $A_0$ across $\mathcal Y^{\mathrm{O}(1)}$ acts as a current whose divergence on $\mathcal R$ cancels that of the D8-brane currents ending there, which keeps \eqref{eq:D8source} and the coupling \eqref{eq:D8coupling} consistent. In the same way, $H_3$ fluxes and NS5-branes are not constrained by the R7-brane, the first nonvanishing RR flux through a cycle carried around it vanishes without sources, and a D8-brane cannot encircle it (Figure~\ref{fig:D8cases}d). If $F_0$ is integrally quantized, the charge ending on an R7-brane must be even; whether the R7-brane shifts the quantization, as orientifold planes do, is a question that the bulk theory does not answer.

The relation \eqref{eq:R7} has a counterpart in type I$'$, where the Romans mass changes sign under the reflection $I_9\Omega$ that defines the O8-planes, and the jump is compensated by D8-brane charge~\cite{Polchinski:1995df,Bergshoeff:2001pv}. In our action, $I_9\Omega$ is the symmetry $-\tau$ of Sec.~\ref{sec:dictionary}. Since $\mathrm O(1)$ is gauged, gauging $\tau$ as well would allow a non-orientable $M_{11}$, with the Poincar\'e dual of $w_1(TM_{11})$ in the role of $\mathcal Y^{\mathrm{O}(1)}$, and would bring in the O8-planes and the defects on which their gluing hypersurfaces end. In M-theory, $(-1)^{F_L}$ and $I_9\Omega$ are both reflections accompanied by $C_3\to-C_3$, of the eleventh circle and of $x^9$ respectively~\cite{Chakrabhavi:2025pintopia}. The circle is therefore reversed around the R7-brane, which suggests that the massless configurations lift to non-orientable backgrounds, on which M-theory requires a $\mathrm{Pin}^+$ structure~\cite{Witten:1996md,Freed:2019sco}. A complete treatment, which requires the fermions, is left to future work.

\section{Algebraic structure}
\label{sec:algebra}

The Gauss laws \eqref{eq:bulk} are the Maurer--Cartan equation of a Lie superalgebra $\mathfrak g$ with ten generators, and $S_{\rm TFT}$ is its Chern--Simons functional. In this section, we describe $\mathfrak g$ (Sec.~\ref{sec:superalgebra}), write the boundary term in terms of a twist and discuss other boundary conditions (Sec.~\ref{sec:twist}), and compare $\mathfrak g$ and the twist with known structures (Sec.~\ref{sec:known}).

\subsection{The gauge superalgebra}
\label{sec:superalgebra}
\label{sec:doubleext}

We associate a generator $e_p$ of degree $1-p$ with each field $A_p$, so that the polyform $\Psi=\sum_pA_p\,e_p$ has total degree one. The RR generators $e_{2k}$ are then odd, and the NSNS generators $e_1$, $e_3$, $e_7$ and $e_9$ are even. With the convention $[A\,e,B\,f]=(-1)^{|e||B|}AB\,[e,f]$, where $|e|$ is the degree of $e$ and $|B|$ the form degree of $B$, the Maurer--Cartan equation ${\rm d}\Psi+\frac12[\Psi,\Psi]=0$ reproduces the Gauss laws \eqref{eq:bulk} for the brackets
\begin{align}
[e_1,e_p]&=-w_p\,e_p\,,\qquad [e_3,e_{2k}]=-e_{2k+2}\,,\nonumber\\
[e_0,e_8]&=-e_7\,,\qquad [e_2,e_6]=e_7\,,\qquad [e_4,e_4]=-e_7\,,\nonumber\\
[e_3,e_7]&=\tfrac12\,e_9\,,\qquad [e_0,e_{10}]=\tfrac54\,e_9\,,\qquad
[e_2,e_8]=-\tfrac34\,e_9\,,\qquad [e_4,e_6]=\tfrac14\,e_9\,,
\label{eq:brackets}
\end{align}
together with those that follow by graded antisymmetry; all others vanish. The graded Jacobi identity is the integrability of the Gauss laws (Sec.~\ref{sec:weights}). Unlike in type IIB, where all the generators are even~\cite{Joung:IIB}, $\mathfrak g$ is a genuine superalgebra. After a rescaling of the generators, it is the superalgebra of the doubled formalism of massive type IIA~\cite{Cremmer:1998px,Lavrinenko:1999xi}, completed by two brackets in the first line of \eqref{eq:brackets}, $[e_1,e_{10}]$ and $[e_3,e_8]$, which the Jacobi identity requires but which are invisible in ten dimensions (Sec.~\ref{sec:known}). What is new here is the use of its invariant pairing to write an eleven-dimensional Chern--Simons action.

The invariant pairing, $\langle[x,y],z\rangle=\langle x,[y,z]\rangle$, is
\begin{equation}
\langle e_1,e_9\rangle=2\,,\qquad \langle e_3,e_7\rangle=1\,,\qquad \langle e_{2k},e_{10-2k}\rangle=(-1)^{k+1}\,.
\label{eq:pairing}
\end{equation}
It is graded symmetric and pairs $e_p$ with $e_{10-p}$. On the odd RR generators, it is antisymmetric: it is the Mukai pairing of the RR fields. Up to normalization, it is the only invariant form that pairs $e_p$ with $e_{10-p}$, as an eleven-dimensional action requires.\footnote{Without this requirement, $\mathfrak g$ has one more invariant graded-symmetric form, $\langle e_1,e_1\rangle$, which exists because $e_1\notin[\mathfrak g,\mathfrak g]$. It is degenerate and pairs $A_1$ with itself in a three-form, so that it cannot enter an eleven-dimensional action.} With the same sign convention as for the bracket, the bulk term \eqref{eq:S11} is
\begin{equation}
S_{\rm TFT}=\frac i2\int_{M_{11}}\langle\Psi,{\rm d}\Psi\rangle+\frac i6\int_{M_{11}}\langle\Psi,[\Psi,\Psi]\rangle\,,
\label{eq:CS}
\end{equation}
that is, $i$ times the Chern--Simons functional of $\mathfrak g$. Given the Gauss laws, the bulk term is therefore unique up to normalization, including the relative coefficients in \eqref{eq:S11}. As in type IIB, $S_{\rm TFT}$ is an AKSZ sigma model~\cite{Alexandrov:1995kv}, whose boundary conditions fit the BV--BFV formalism~\cite{Cattaneo:2012qu}.

The gauge transformations of \eqref{eq:CS}, $\delta\Psi={\rm d}\Lambda+[\Psi,\Lambda]$ with $\Lambda=\sum_p\varepsilon_{p-1}\,e_p$, are the $\mathbb{R}^+$ transformations \eqref{eq:Rgauge}, with parameter $\varepsilon_0$, and the higher-form gauge transformations \eqref{eq:hfgauge}. The element $-1\in\mathrm{O}(1)$ reverses the odd generators: $(-1)^{F_L}$ is the $\mathbb{Z}_2$ grading of $\mathfrak g$, and the global form of the gauge group enters only through it.

The brackets \eqref{eq:brackets} are not independent. We split the generators as
\begin{equation}
\mathfrak g=\mathfrak b\oplus V\oplus\mathfrak b^*\,,\qquad
\mathfrak b={\rm span}\{e_1,e_3\}\,,\quad V={\rm span}\{e_0,e_2,\dots,e_{10}\}\,,\quad \mathfrak b^*={\rm span}\{e_9,e_7\}\,.
\label{eq:dext}
\end{equation}
Here $\mathfrak b$ is a Lie algebra, $[e_1,e_3]=e_3$; $V$ is spanned by the odd RR generators and carries the Mukai pairing; and $\mathfrak b^*$ is dual to $\mathfrak b$ under \eqref{eq:pairing}. The independent data are the action of $\mathfrak b$ on $V$, given by the first line of \eqref{eq:brackets}, and the Mukai pairing, which this action preserves. The generator $e_1$ multiplies $e_p$ by $-w_p$, and $e_3$ maps each RR generator to the next one along the chain
\begin{equation}
e_0\to e_2\to e_4\to e_6\to e_8\to e_{10}\to0\,.
\label{eq:chain}
\end{equation}
All other brackets follow from the invariance of the pairing. Those of $\mathfrak b$ with $\mathfrak b^*$ are the coadjoint action, $\langle[x,\xi],y\rangle=-\langle\xi,[x,y]\rangle$ for $x,y\in\mathfrak b$ and $\xi\in\mathfrak b^*$:
\begin{equation}
[e_1,e_7]=-e_7\,,\qquad [e_3,e_7]=\tfrac12\,e_9\,,\qquad [e_1,e_9]=[e_3,e_9]=0\,.
\label{eq:coadj}
\end{equation}
Those of two RR generators have no component along $V$, so that $V$ is abelian, and take values in $\mathfrak b^*$:
\begin{equation}
[v,v']=\tfrac12\,\langle v,[v',e_1]\rangle\,e_9+\langle v,[v',e_3]\rangle\,e_7\,,\qquad v,v'\in V\,.
\label{eq:VV}
\end{equation}
Together, \eqref{eq:coadj} and \eqref{eq:VV} give the last two lines of \eqref{eq:brackets}: the RR terms in the Gauss laws of $A_7$ and $A_9$ follow from the first line. The superalgebra $\mathfrak g$ is therefore determined by the $\mathfrak b$-module $V$ and its invariant pairing. In the terminology of Medina and Revoy, it is the double extension of $V$ by $\mathfrak b$~\cite{Medina:1985,FigueroaOFarrill:1995opp,Benamor:1999}; in particular, it is quadratic, as a Chern--Simons functional requires.

The weights enter only through the conditions \eqref{eq:neutral}, which are those for $\mathfrak b$ to act on $V$ and preserve the pairing. These conditions fix the weights up to an overall factor, the normalization of $e_1$. Since $[e_1,e_3]=e_3$, each step of \eqref{eq:chain} lowers the weight by one, and the invariance of $\langle e_0,e_{10}\rangle$ requires $w_0+w_{10}=0$. Hence $w_{2k}=\frac{5-2k}{2}$. In massless type IIA, the chain runs from $e_2$ to $e_8$. The generators $e_0$ and $e_{10}$ extend it at its two ends. The new step, $[e_3,e_0]=-e_2$, is the St\"uckelberg coupling, the algebraic counterpart of the fundamental strings created when a D0-brane crosses a D8-brane~\cite{Danielsson:1997wq,Bachas:1997kn}. Since no weight vanishes, $V$ splits into its parts of positive and negative weight, $V=V_+\oplus V_-$, with $V_+={\rm span}\{e_0,e_2,e_4\}$ and $V_-={\rm span}\{e_6,e_8,e_{10}\}$. They hold the RR field strengths $F_0$, $F_2$ and $F_4$ of Romans' theory and their duals, as in the democratic formulation~\cite{Bergshoeff:2001pv}. The Mukai pairing relates $V_+$ to $V_-$, and $e_3$ preserves $V_-$ but not $V_+$, since it maps $e_4$ to $e_6$.

The structure of $\mathfrak g$ also explains why the reduction of Sec.~\ref{sec:reduction} works. The magnetic fields are the fields along $\mathfrak a=V_-\oplus\mathfrak b^*={\rm span}\{e_6,\dots,e_{10}\}$. This subspace is Lagrangian, i.e. it contains exactly one generator of each pair. It is also abelian, and it is an ideal. Since $\mathfrak a$ is Lagrangian and abelian, the bulk term is linear in the magnetic fields, which therefore act as Lagrange multipliers, as in \eqref{eq:SBF}. Since $\mathfrak a$ is an ideal, the constraints they impose, \eqref{eq:constraints}, are the flatness conditions of the quotient $\mathfrak g/\mathfrak a\cong\mathfrak b\ltimes V_+$. This is the algebra of Romans' potentials: $e_1$ acts by the weights, and $e_3$ along $e_0\to e_2\to e_4$. The constraints are therefore solved locally by pure-gauge configurations, whose gauge parameters are the potentials of Romans' theory (Secs.~\ref{sec:solve} and~\ref{sec:gaugedata}).

Conversely, $\mathfrak g$ is determined by $\mathfrak b\ltimes V_+$ and one three-form on it. The pairing identifies $\mathfrak a$ with the dual of $\mathfrak b\ltimes V_+$, so that $\mathfrak g$ consists of $\mathfrak b\ltimes V_+$ and its dual, paired with each other. Besides the brackets of $\mathfrak b\ltimes V_+$ and its coadjoint action on the dual, $\mathfrak g$ has only $[e_3,e_4]=-e_6$ and $[e_4,e_4]=-e_7$, which take values in the dual. Both come from the three-form $\langle[e_3,e_4],e_4\rangle$ on $\mathfrak b\ltimes V_+$, which is the cubic term $A_3A_4A_4$ of \eqref{eq:Sred}. In the language of Refs.~\cite{Bordemann:1997,Bajo:2000}, $\mathfrak g$ is the $T^*$-extension of the algebra of Romans' potentials by its Chern--Simons term. The double extension above builds $\mathfrak g$ from its RR part $V$; the $T^*$-extension builds it from the potentials of Romans' theory.

Finally, $\mathfrak g$ has two scaling symmetries. Every bracket has the form $[e_p,e_q]\propto e_{p+q-1}$, so that a rescaling $e_p\to\Omega^{d_p}e_p$ preserves all brackets if $d_p+d_q=d_{p+q-1}$. This condition has two independent solutions,
\begin{equation}
d_p=w_p\,,\qquad d_p=p-1\,,
\label{eq:scalings}
\end{equation}
which correspond to the two scaling symmetries of type IIA (Sec.~\ref{sec:dictionary}). The first is generated by $e_1$ and preserves the pairing: it is the $\mathbb{R}^+$ gauge symmetry. The second is not generated by any element of $\mathfrak g$, and it multiplies the pairing, and hence $S_{\rm TFT}$, by $\Omega^8$, since $e_p$ and $e_{10-p}$ have total degree eight. In type IIA, it is the scaling $g\to\Omega^2g$, $F_p\to\Omega^{p-1}F_p$, which also multiplies the action by $\Omega^8$: it maps solutions to solutions, but it is not a symmetry of the action. It therefore has no counterpart among the gauge symmetries of $S_{\rm TFT}$.

\subsection{The twist and other boundary conditions}
\label{sec:twist}

The Chern--Simons functional \eqref{eq:CS} is a bulk term only. The boundary term \eqref{eq:S10} can be written in the same language, but it needs one more piece of data. The Hodge star maps a $p$-form to a $(10-p)$-form, whereas the pairing relates $e_p$ to $e_{10-p}$, so that $\langle\Psi,*\Psi\rangle$ contains no ten-form. A linear map $\mathcal S$ on $\mathfrak g$ is needed to compensate, and we call it the twist. With the twist
\begin{equation}
\mathcal S\,e_1=2\,e_9\,,\qquad \mathcal S\,e_9=\tfrac12\,e_1\,,\qquad \mathcal S\,e_3=e_7\,,\qquad \mathcal S\,e_7=e_3\,,\qquad \mathcal S\,e_{2k}=(-1)^{k+1}\,e_{10-2k}\,,
\label{eq:twist}
\end{equation}
and $*\mathcal S\,\Psi=\sum_p(*A_p)\,\mathcal S\,e_p$, the boundary term \eqref{eq:S10} and the boundary conditions \eqref{eq:bcproj} become
\begin{equation}
S_{\rm b}=\frac14\int_{M_{10}}\langle\Psi,*\mathcal S\,\Psi\rangle\,,\qquad\qquad *\mathcal S\,\Psi=-i\,\Psi\quad\text{on }M_{10}\,,
\label{eq:Sbtwist}
\end{equation}
with $*\mathcal S\,\Psi=\Psi$ in Lorentzian signature. Together with \eqref{eq:CS}, this gives $S_{\rm bTFT}$ in terms of $\Psi$ alone.

The twist is strongly constrained. First, the integrand of \eqref{eq:Sbtwist} must be a ten-form. Since $\mathfrak g$ has one generator in each degree, this requires $\mathcal S\,e_p=s_p\,e_{10-p}$. Second, the condition $*\mathcal S\,\Psi=-i\,\Psi$ contains ten equations, one for each $A_p$, and the two equations of a pair must agree, so that the condition selects half of the boundary values rather than none. This requires $(*\mathcal S)^2=-1$ (and $+1$ in Lorentzian signature), that is, $s_p\,s_{10-p}=1$ for odd $p$ and $-1$ for even $p$; it is the condition on the coefficients of Sec.~\ref{sec:bdry}. Equivalently, $\mathcal S^2=+1$ on the NSNS generators and $-1$ on the RR generators: $\mathcal S^2$ is the parity automorphism of $\mathfrak g$, the action of $-1\in\mathrm{O}(1)$ (Sec.~\ref{sec:superalgebra}), and $(*\mathcal S)^2=*^2\,\mathcal S^2=-1$ because $*^2$ is $-1$ on odd forms and $+1$ on even forms, opposite to $\mathcal S^2$. Third, the kinetic terms of the reduced action \eqref{eq:Sred} are positive only if $s_p\,\langle e_p,e_{10-p}\rangle>0$. The twist is therefore fixed up to one positive number for each of the five pairs: it exchanges each generator with its partner, with fixed signs. A constant $\mathbb{R}^+$ transformation and a rescaling of the metric remove two of these numbers (Sec.~\ref{sec:bdry}). The remaining three are couplings of the boundary theory, and \eqref{eq:twist} sets them to their type IIA values. The superalgebra $\mathfrak g$ does not fix these three couplings; in type IIA, supersymmetry does.

The twist determines the unbroken gauge symmetry. A constant gauge transformation preserves $S_{\rm b}$ only if it commutes with $\mathcal S$. Since $w_{10-p}=-w_p$, the twist anticommutes with the action of $e_1$, and an element $\lambda$ of $\mathrm{GL}(1,\mathbb{R})$ commutes with it only if $\lambda^{4w_p}=1$, that is, for $\lambda=\pm1$. The breaking of $\mathrm{GL}(1,\mathbb{R})$ to $\mathrm{O}(1)$ thus follows from the form of the twist, and does not depend on the three couplings.

In our setup, the twisted condition on $M_{10}=\partial M_{11}$ fixes half of the boundary values by a twist map, which depends on the metric (in type IIB, also by a projection). Any boundary condition selects a Lagrangian subspace of the boundary values, with respect to the symplectic form defined by \eqref{eq:theta}. A boundary of $M_{11}$ could instead carry a topological condition, which makes this choice by the algebra alone: a Dirichlet condition fixes the boundary value of one field of each pair and leaves that of its partner free. Let $\mathfrak d$ be the span of the generators whose boundary values are fixed, and $\mathfrak d^*$ that of their partners, so that $\mathfrak g=\mathfrak d\oplus\mathfrak d^*$. Then $\mathfrak d^*$ contains exactly one generator of each pair, that is, it is Lagrangian, $\mathfrak d^*=\mathfrak l$ with $\mathfrak l^\perp=\mathfrak l$, where $^\perp$ is the orthogonal complement with respect to the pairing \eqref{eq:pairing}. For the Dirichlet condition to be gauge invariant, $\mathfrak l$ must also be a subalgebra, as in Chern--Simons theory~\cite{Kapustin:2010hk,Severa:2016lwc,Pulmann:2019vrw}. The condition then leaves no dynamics on the boundary. Of the $2^5$ graded Lagrangian subspaces, eighteen are subalgebras: the sixteen that contain $e_7$, and the two that contain $e_3$, $e_6$, $e_8$ and $e_{10}$. Only one of them is an ideal, the $\mathfrak a$ that underlies the reduction (Sec.~\ref{sec:superalgebra}).

In the language of the SymTFT~\cite{Freed:2012bs,Gaiotto:2020iye,Apruzzi:2021nmk,Freed:2022qnc}, $M_{10}$ with the twisted condition is a physical boundary, and a boundary with a topological condition is a symmetry boundary. Unlike in most SymTFT constructions, where the physical boundary supports a relative field theory that is specified separately, here the physical boundary condition follows from the local boundary term $S_{\rm b}$, and the boundary theory, including its action, is obtained by reduction. The two meet in the sandwich construction, in which $M_{11}=M_{10}\times[0,1]$ has the physical boundary at one end and a symmetry boundary at the other. Shrinking the interval leaves a theory on $M_{10}$ in which the fields along $\mathfrak d$ are fixed backgrounds and those along $\mathfrak l$ are summed over, so that different choices of $\mathfrak l$ give dual formulations, as in Poisson--Lie T-duality~\cite{Severa:2016lwc}. For the pair $(A_0,A_{10})$, $e_{10}\in\mathfrak l$ fixes $A_0$ and makes $m$ a parameter, as in supergravity, whereas $e_0\in\mathfrak l$ fixes $A_{10}$ and sums over $m$, giving the nine-form formulation, in which D8-branes are electric sources. These are the two symmetry boundaries of the SymTFT of a $(-1)$-form symmetry~\cite{Najjar:2024minus1} (Sec.~\ref{sec:discussion}). In the reduction of Sec.~\ref{sec:reduction}, which has no symmetry boundary, $m$ is instead varied (Sec.~\ref{sec:zeromode}).

A boundary cannot carry both the twisted condition and a topological condition, since each already fixes half of the boundary values. The two can, however, be imposed on different pairs: the Dirichlet condition on some of them, which are dropped from $S_{\rm b}$, and the twisted condition on the others. For example, the Dirichlet condition $A_0=0$ on the pair $(A_0,A_{10})$, with the twisted condition on the other pairs, gives massless type IIA (Sec.~\ref{sec:zeromode}). In general, $\mathfrak d\oplus\mathfrak d^*$ is a proper subspace of $\mathfrak g$, and the boundary value of $\Psi$ lies in $\mathfrak c\cong\mathfrak g/\mathfrak d$, the span of all generators not in $\mathfrak d$. This subspace is coisotropic, $\mathfrak c^\perp=\mathfrak d^*\subset\mathfrak c$, and the twisted condition is imposed on $\mathfrak c/\mathfrak c^\perp\cong\mathfrak g/(\mathfrak d\oplus\mathfrak d^*)$, as in the reduction of Courant algebroids~\cite{Bursztyn:2005rbc}. As before, $\mathfrak d^*$ must be a subalgebra. For massless type IIA, $\mathfrak d^*={\rm span}\{e_{10}\}$. In type IIB~\cite{Joung:IIB}, $\mathfrak d^*=\mathfrak{so}(2)$, and the Dirichlet condition leaves the boundary value of the $\mathfrak{so}(2)$ part of $A_1$ free. For $\mathfrak d=0$, no pair carries the Dirichlet condition, and this reduces to the twisted condition of this paper. For $\mathfrak d\oplus\mathfrak d^*=\mathfrak g$, every pair carries it, none is left for the twist, and the boundary is a symmetry boundary with $\mathfrak l=\mathfrak d^*$.

\subsection{Relation to known structures}
\label{sec:known}

In ten dimensions, the algebra of the type IIA forms has been described on shell in several ways. We compare these descriptions with $\mathfrak g$, starting from the Bianchi identities, and end with the twisted condition.

On $M_{10}$, the Maurer--Cartan equation of $\mathfrak g$ reduces to the Bianchi identities \eqref{eq:bianchi}. The Chevalley--Eilenberg algebra of $\mathfrak g$ is therefore the free differential algebra of democratic massive type IIA~\cite{Bergshoeff:2001pv,Bergshoeff:2006qw}, in the $L_\infty$ language of Refs.~\cite{Sati:2008eg,Fiorenza:2013nha,Giotopoulos:2026cqts}, up to the eleven-form relation discussed below, with the dilaton included through $A_1$ and $A_9$.

In the doubled formalism of Refs.~\cite{Cremmer:1998px,Lavrinenko:1999xi} (see also Ref.~\cite{Rosabal:2025}), the field strengths are collected in a superalgebra-valued form $\mathcal G$, in which each potential has a dual partner, and the field equations are imposed as the twisted self-duality constraint $*\mathcal G=\mathcal S\,\mathcal G$. For massive type IIA, Ref.~\cite{Lavrinenko:1999xi} gives
\begin{equation}
\mathcal G=\tfrac12\,{\rm d}\phi\,H+m\,e^{\frac54\phi}\,Y+e^{\frac34\phi}F_2\,W_1+e^{-\frac12\phi}H_3\,V^1+\cdots\,.
\label{eq:doubled}
\end{equation}
On the solution \eqref{eq:dict}, with ${\cal V}=e^{\phi/2}$, this coincides with $\Psi$. In our action, the Maurer--Cartan equation of $\mathcal G$ is the set of bulk Gauss laws, and the constraint $*\mathcal G=\mathcal S\,\mathcal G$ is the boundary condition \eqref{eq:bcproj}. The twist is not an automorphism of $\mathfrak g$: for instance, $\mathcal S[e_1,e_3]=e_7$, whereas $[\mathcal S e_1,\mathcal S e_3]=2[e_9,e_7]=0$.

After a rescaling of the generators, the brackets of Ref.~\cite{Lavrinenko:1999xi} are brackets of $\mathfrak g$, with $(e_0,e_1,e_2,e_3,e_4)\leftrightarrow(Y,H,W_1,V^1,V)$ and $(e_6,e_7,e_8,e_9,e_{10})\leftrightarrow(\widetilde V,\widetilde V_1,\widetilde W^1,\widetilde H,\widetilde Y)$. The superalgebra $\mathfrak g$ has two more brackets, both in the first line of \eqref{eq:brackets}: $[e_1,e_{10}]$, which gives $A_{10}$ its $\mathbb R^+$ weight, and $[e_3,e_8]=-e_{10}$, the last step of \eqref{eq:chain}. Both enter only the Gauss law ${\rm D}A_{10}=A_3A_8$, an eleven-form, and are therefore invisible in ten dimensions. Ref.~\cite{Lavrinenko:1999xi} does not give the first and includes the second only above ten dimensions, but the Jacobi identity requires both, with the coefficients of \eqref{eq:brackets}. It also uses no invariant pairing, which our action requires. In Ref.~\cite{Cremmer:1998px}, the massless superalgebra is presented as a deformation of $G\ltimes G^*$, with $G={\rm span}\{e_1,\dots,e_4\}$ and $G^*={\rm span}\{e_6,\dots,e_9\}$; this splitting differs from \eqref{eq:dext}.

The relation missing from the free differential algebra, mentioned above, is the Gauss law of $A_{10}$: its Maurer--Cartan component, ${\rm d}F_{10}=H_3F_8$, is an eleven-form. In ten dimensions, the bracket $[e_3,e_8]$ appears only in the gauge transformation of the nine-form~\cite{Lavrinenko:1999xi,Bergshoeff:2006qw}. Reference~\cite{Lavrinenko:1999xi} obtained it in this way, or by formally extending the Bianchi identities to eleven dimensions, and used it to derive the tension relation $T_{\rm D8}=T_{\rm F1}T_{\rm D6}/2\pi$ without T-duality. In our action, this Gauss law is the equation of motion of $A_0$ (Sec.~\ref{sec:zeromode}), and it involves no new input. The cubic term $A_3A_0A_8$ of $S_{\rm TFT}$ appears in three Gauss laws, one for each of its fields:
\begin{equation}
{\rm D} A_2=A_3A_0\,,\qquad {\rm D} A_7=A_0A_8+\cdots\,,\qquad {\rm D} A_{10}=A_3A_8\,.
\label{eq:threelaws}
\end{equation}
On $M_{10}$, the first two are the Bianchi identities ${\rm d} F_2=F_0H_3$ and ${\rm d} H_7=F_0F_8+\cdots$ of massive type IIA, and they fix the third. In the abelian eleven-dimensional Chern--Simons theory of Ref.~\cite{Belov:2006xj}, the corresponding term is a twist of the differential by a background $H_3$; here, it is a bracket of the gauge algebra, and $H_3$ and $H_7$ are dynamical.

Borcherds superalgebras~\cite{HenryLabordere:2002dk,Greitz:2011vh} and tensor hierarchy algebras~\cite{Palmkvist:2013vya} organize the potentials rather than the field strengths. They place all the potentials and their duals in one algebra, graded by the form degree $q$. In $\mathfrak g$, the generator $e_p$ corresponds to a potential of degree $q=p-1$, so that $-1\le q\le9$. The Borcherds superalgebra of type IIA (Sec.~4.2 of Ref.~\cite{Greitz:2011vh}) is generated by two simple roots, an odd one for $C_1$ and an even one for $B_2$, which correspond to $e_2$ and $e_3$. Its generators for $H_7$ and for the dual of the dilaton arise from repeated brackets of these, just as $[e_4,e_4]$ and $[e_3,[e_4,e_4]]$ do in $\mathfrak g$. For $q>0$, the two algebras therefore appear to agree. The Borcherds superalgebra, however, is infinite-dimensional, and is cut off at $q=9$. In $\mathfrak g$, no cut-off is needed: the pairing gives each $e_p$ a partner $e_{10-p}$, so that $\mathfrak g$ has only ten generators. In addition, $\mathfrak g$ has a generator at $q=-1$, namely $e_0$. In tensor hierarchy algebras, this degree holds the embedding tensor, a deformation parameter; in $\mathfrak g$, it holds the Romans mass, as the value of the field $A_0$. In both kinds of algebras, the potentials of degrees $q$ and $8-q$ are dual to each other; in $\mathfrak g$, this duality is the pairing of $e_p$ with $e_{10-p}$.

The twisted condition, which is the twisted self-duality constraint of the doubled formalism~\cite{Cremmer:1998px,Lavrinenko:1999xi}, also has counterparts in Chern--Simons theory and in linearized gravity. It is of the form $*A=EA$, with $E$ a generalized metric on the gauge algebra, which is the non-topological boundary condition of Chern--Simons theory in Ref.~\cite{Severa:2016lwc} and of its higher analogues~\cite{Pulmann:2019vrw}. A boundary term and a boundary condition of the form \eqref{eq:Sbtwist} also appear in the description of four-dimensional linearized gravity as edge modes of a five-dimensional topological theory~\cite{Chen:2026edge}. There, the field takes values in the three-forms of $\mathbb{R}^{2,4}$, and the Hodge star of $\mathbb{R}^{2,4}$ plays the role of the twist. This map commutes with $\mathfrak{so}(2,4)$, so that the bilinear form of the boundary term is invariant under the whole algebra, like that of the bulk term. Here, the bilinear form of the boundary term is invariant only under $\mathrm{O}(1)$, and the broken $\mathbb{R}^+$ gives the dilaton.

\section{Discussion}
\label{sec:discussion}

We have constructed an eleven-dimensional topological action $S_{\rm bTFT}=S_{\rm TFT}+S_{\rm b}$ for the dilaton and form-field sector of massive type IIA. The bulk term is fixed by the gauge superalgebra and its invariant pairing (Sec.~\ref{sec:algebra}). The gauge superalgebra is that of the doubled formalism~\cite{Cremmer:1998px,Lavrinenko:1999xi}, completed by two brackets that are invisible in ten dimensions and give the Gauss law of degree eleven, the equation of motion of the Romans mass. The boundary term, the only place where the metric enters, breaks $\mathrm{GL}(1,\mathbb{R})$ to $\mathrm{O}(1)$, and the reduction to the boundary gives Romans' action off shell, with the dilaton arising as an edge mode (Sec.~\ref{sec:romans}). Globally, the holonomies of the flat connection are monodromies of the dilaton and of $(-1)^{F_L}$, and for nonzero Romans mass they are related to the D8-brane charges; the hypersurface dual to the $\mathrm O(1)$ holonomy can end on reflection seven-branes (Sec.~\ref{sec:global}).

The construction follows that of type IIB~\cite{Joung:IIB}, with $\mathrm{GL}(1,\mathbb{R})\supset\mathrm{O}(1)$ in place of $\mathrm{SL}(2,\mathbb{R})\supset\mathrm{SO}(2)$, and differs from it in three respects. First, no field is paired with itself, so that the reduction gives an action rather than a pseudo-action. Second, the unbroken group is discrete, so that, unlike in type IIB, no Dirichlet condition is needed. Third, the zero-form $A_0$ makes the Romans mass a bulk field: its Gauss law is an eleven-form, whose integral is the equation of motion of $m$ (Sec.~\ref{sec:zeromode}). {The general structure, however, is the same}. The boundary terms \eqref{eq:theta} give a symplectic form on the space of boundary values, the conditions \eqref{eq:bcproj} select a Lagrangian subspace, with $S_{\rm b}$ as its generating function, and other Lagrangian subspaces give other boundary theories for the same bulk, as in the SymTFT framework~\cite{Kapustin:2010hk,Maldacena:2001ss,Freed:2012bs,Gaiotto:2020iye,Apruzzi:2021nmk,Heckman:2022xgu,Apruzzi:2023uma,Bonetti:2024symtft,Antinucci:2024ssb,Apruzzi:2025spacetime,Bonetti:2025nonlinear} (Sec.~\ref{sec:twist}), with the difference that the physical boundary is fixed here by a local boundary term rather than by a separately given boundary theory. In this language, the pair $(A_0,A_{10})$ has the form of the SymTFT of a $(-1)$-form symmetry, whose parameter is $m$~\cite{Najjar:2024minus1}; in holography, the Romans mass has been interpreted as the background of a $(-2)$-form symmetry of a three-dimensional dual~\cite{Banerjee:2026minus2}.

So far, our analysis has been classical, and the most immediate next step is to quantize the theory, or at least to examine its quantum aspects. A first such aspect is the role of the bulk: the boundary theory obtained by the reduction still depends on the bulk configuration, through its Chern--Simons terms and holonomies (Secs.~\ref{sec:fromboundary} and~\ref{sec:holonomy}), and requiring the quantum bulk theory to be consistent, so that this dependence drops out of the boundary physics, should lead to quantization conditions. For the RR fluxes, these are expected to be those of twisted K-theory~\cite{Minasian:1997mm,Witten:1998cd,Moore:1999gb}, with the boundary condition chosen accordingly, as in the eleven-dimensional Chern--Simons theory of Ref.~\cite{Belov:2006xj}. Flux quantization should also restrict the holonomy. An $\mathbb{R}^+$ holonomy rescales the charges of a D-brane and of its magnetic dual by inverse factors, which is incompatible with integral charges. Only the $(-1)^{F_L}$ monodromy should then survive, as only $\mathrm{SL}(2,\mathbb{Z})$ monodromies do in type IIB, which leaves only $W_0(\gamma)=\pm1$ in \eqref{eq:D8hol} (Sec.~\ref{sec:D8}).

Several further directions seem natural. The first is the relation to type IIB in nine dimensions. On a circle, massive type IIA is T-dual to type IIB with a parabolic $\mathrm{SL}(2,\mathbb{R})$ monodromy~\cite{Bergshoeff:1996ui,Hull:1998vy}, and the Romans mass, which here enters through $A_0$, becomes there a holonomy of the flat $\mathrm{SL}(2,\mathbb{R})$ connection of Ref.~\cite{Joung:IIB}. Since T-duality mixes the form fields with the metric, relating the two bulk theories requires including gravity in the bulk. Such an extension would also be the natural setting in which to revisit the one-loop coupling $B_2\wedge X_8$, which involves the curvature and breaks the $\mathbb{R}^+$ symmetry (Sec.~\ref{sec:democratic}). The second is gauged supergravity. The pair $(A_0,A_{10})$ suggests how its deformation parameters could enter the construction: as zero-forms in a representation of the duality group, paired with top forms, at the degree where tensor hierarchy algebras place the embedding tensor (Sec.~\ref{sec:known}). The scalar potential produced by the gauging would then arise as the potential term of Romans' theory does here. We treat lower-dimensional maximal supergravities in a forthcoming paper~\cite{Joung:lower}. Finally, gauging also the orientation-reversing $\mathbb Z_2$ of \eqref{eq:orrev} would admit a non-orientable $M_{11}$, and with it the O8-planes of type I$'$ and the defects on which the corresponding gluing hypersurfaces end (Secs.~\ref{sec:dictionary} and~\ref{sec:R7}). We return to these questions, and to branes in the eleven-dimensional bulk in general, in a forthcoming paper~\cite{Joung:inprep}.

\appendix
\section{Conventions and Lorentzian signature}
\label{app:conventions}

Wedge products are implicit, and the exterior derivative acts from the left, ${\rm d}(\alpha\beta)={\rm d}\alpha\,\beta+(-1)^{p}\alpha\,{\rm d}\beta$ for a $p$-form $\alpha$. The boundary $M_{10}=\partial M_{11}$ is oriented so that Stokes' theorem reads $\int_{M_{11}}{\rm d}\omega=\int_{M_{10}}\omega$. The Hodge star of $M_{10}$ is defined by $\alpha*\beta=(\alpha,\beta)\,{\rm vol}$ for two $p$-forms, with $(\alpha,\beta)=\frac1{p!}\,\alpha_{\mu_1\cdots\mu_p}\beta^{\mu_1\cdots\mu_p}$ and ${\rm vol}$ the volume form of the metric. Then $\alpha*\beta=\beta*\alpha$, and on $p$-forms $*^2=(-1)^p$ in Euclidean signature and $*^2=(-1)^{p+1}$ in Lorentzian signature.

In Euclidean signature, with the weight $e^{-S_{\rm bTFT}}$ and real fields, $S_{\rm b}$ is positive definite, while $S_{\rm TFT}$ is imaginary. The boundary conditions \eqref{eq:bcproj} are then complex: for real electric fields, the boundary values \eqref{eq:magnetic} of the magnetic fields are imaginary. They are the saddle point of the Gaussian integral over the real boundary values of the magnetic fields. This is the sense in which the fields are complexified.

In Lorentzian signature, with the weight $e^{iS_{\rm bTFT}}$, the action is
\begin{equation}
S_{\rm bTFT}=\frac12\int_{M_{11}}\Big\{\cdots\Big\}-\frac14\int_{M_{10}}\Big\{\cdots\Big\}\,,
\label{eq:SLor}
\end{equation}
where the two brackets are those of \eqref{eq:S11} and \eqref{eq:S10}. The decomposition \eqref{eq:SBF}--\eqref{eq:Sdel} holds with $i$ replaced by $1$, and the boundary conditions are
\begin{equation}
*A_1=\tfrac12\,A_9\,,\qquad *A_3=A_7\,,\qquad *A_{2k}=(-1)^{k+1}A_{10-2k}\,.
\label{eq:bcLor}
\end{equation}
On them, the boundary term of \eqref{eq:SLor} vanishes, and the reduced action is
\begin{equation}
S_{\rm bTFT}=-\int_{M_{10}}\Big[2\,A_1*A_1+\tfrac{1}{2}\,A_3*A_3+\tfrac{1}{2}\sum_{k=0}^{2}A_{2k}*A_{2k}\Big]-\tfrac{1}{2}\int_{M_{11}}A_3A_4A_4\,.
\label{eq:SredLor}
\end{equation}
On the solution \eqref{eq:puregauge}, \eqref{eq:HFsol}, this is the Lorentzian form of \eqref{eq:SR}, with the kinetic terms $-\frac12\,{\rm d}\phi*{\rm d}\phi$ and $-\frac12\,e^{w_p\phi}F_p*F_p$, and the Chern--Simons terms $-\frac12\big[B_2\,{\rm d} C_3\,{\rm d} C_3+\frac m3\,B_2^3\,{\rm d} C_3+\frac{m^2}{20}\,B_2^5\big]$.

\acknowledgments

This work was supported by the Brain Pool program (RS-2025-25457100) and the Individual Basic Research program (RS-2025-00564305) of the National Research Foundation of Korea (NRF), funded by the Ministry of Science and ICT (MSIT).

\bibliographystyle{JHEP}
\bibliography{IIA}

\end{document}